\documentclass[fleqn,usenatbib]{mnras}

\usepackage{newtxtext,newtxmath}
\usepackage[T1]{fontenc}

\DeclareRobustCommand{\VAN}[3]{#2}
\let\VANthebibliography\thebibliography
\def\thebibliography{\DeclareRobustCommand{\VAN}[3]{##3}\VANthebibliography}

\usepackage{graphicx}	% Including figure files
\usepackage{amsmath}	% Advanced maths commands

\newcommand{\refbf}{} % use {} to switch off the bf changes from resubmission

\title[6dFGS BELAGN catalogue]{Broad-line Active Galactic Nuclei in the 6dF Galaxy Survey}

\author[W. Hon]{
Wei Jeat Hon,$^{1}$\thanks{E-mail: whon@student.unimelb.edu.au}
Rachel L. Webster,$^{1}$
Christian Wolf$^{2,3}$
\\
$^1$School of Physics, University of Melbourne, Parkville, Victoria 3010, Australia \\
$^2$Research School of Astronomy and Astrophysics (RSAA), The Australian National University, Canberra, ACT 2611, Australia\\
$^3$Centre for Gravitational Astrophysics, Research School of Physics and RSAA, Australian National University, Canberra, ACT 2600, Australia \\
}

\date{Accepted XXX. Received YYY; in original form ZZZ}

\pubyear{2023}

\begin{document}
\label{firstpage}
\pagerange{\pageref{firstpage}--\pageref{lastpage}}
\maketitle

% Abstract of the paper
\begin{abstract}
The Six-degree Field Galaxy Survey (6dFGS) is a spectroscopic redshift survey {\refbf of the Southern hemisphere} completed in 2006. While it provides 136,304 spectra of mostly low-redshift galaxies, a large and reliable catalogue of Active Galactic Nuclei (AGN) that are selected {\refbf based on spectral signatures} is still lacking. In this work, {\refbf we present an extensive list of verified broad emission-line AGN in the 6dFGS sample. We visually confirm the AGN nature of all spectra, and} disentangle fibre cross-talk to remove bogus AGN. The final catalogue contains 2,515 unique broad-line AGN {\refbf with a median redshift of 0.207}, of which 891 are identified for the first time. A flux-limited subsample contains 665 AGN to a K-band magnitude of 13. This new sample adds to the {\refbf list of known low-luminosity} AGN in the Southern hemisphere and thus provides a basis for investigations of low-redshift AGN with the upcoming Vera C. Rubin Observatory. 
\end{abstract}

% Select between one and six entries from the list of approved keywords.
% Don't make up new ones.

%%%%%%%%%%%%%%%%%%%%%%%%%%%%%%%%%%%%%%%%%%%%%%%%%%

%%%%%%%%%%%%%%%%% BODY OF PAPER %%%%%%%%%%%%%%%%%%

\section{Introduction}
% AGN lists and catalogue and their scientific contribution
{\refbf 
Since the initial discoveries of Active Galactic Nuclei \citep[AGN,][]{seyfert43, schmidt63} around a million have been identified, mostly in large surveys, the largest being the Sloan Digital Sky Survey \citep[SDSS,][]{almeida23}. The luminosity of AGN is tightly related to their accretion rate. While the observed UV-optical continuum luminosity also depends on whether or not an AGN is obscured by dust (type-2 vs. type-1), the strength of the narrow [O{\sc iii}]$\lambda\lambda4959,5007$ emission-line doublet is a good indicator of intrinsic AGN luminosity that is almost independent of viewing angle \citep{Kauffmann03} and well-correlated with X-ray luminosity \citep{Heckman05,Ueda15}.  Strong star formation tends to contribute much less to [O{\sc iii}] than to other emission lines.

The luminosity function of AGN is steep \citep[e.g.,][]{Richards06}, so those with low accretion rates, often labelled Seyfert galaxies, vastly outnumber the more luminous, more rapidly accreting objects, often called quasars. AGN emission is also intrinsically variable \citep{Lawrence16}, and it is firmly established that low-luminosity objects show both stronger stochastic variability amplitudes \citep[e.g.][]{VB04,TWT23,Arevalo24} and more common extreme changes, exemplified by Changing-Look AGN \citep[CLAGN;][]{macleod16,RT23}. It has been proposed that this is related to the shorter physical timescales in the smaller accretion discs of lower-luminosity objects \citep{FKR02,TWT23}. For these reasons, low-luminosity AGN are attractive targets for studies of variability mechanisms in general, and for studies of CLAGN and their host galaxies in particular.

It is difficult to identify low-luminosity AGN in imaging surveys as they are often extended with less distinguished colours. Their identification relies on spectroscopic surveys and an analysis dedicated to measuring key emission lines. The first large sample of low-luminosity AGN was identified in SDSS \citep[see e.g.][]{Kauffmann03,GrHo07}. However, the sky coverage of SDSS is restricted to a large part of the Northern sky. In contrast, the Southern sky is lacking comparable AGN catalogues, with the primary bright catalogue covering less than 4\% of the hemisphere \citep[2dF QSO Redshift Survey,][]{Croom04}.

Long-term wide-area optical monitoring of the Southern sky by the Legacy Survey of Space and Time \citep[LSST,][]{ivezic19} at the Vera C. Rubin Observatory will commence in 2025. LSST will provide light curves for every galaxy and AGN in the Southern sky over a ten-year period. These light curves are taken at around a fortnightly cadence for sources down to $i\approx23.5$ (5$\sigma$-limit). This implies that by the end of the survey period, there will be a light curve for every AGN in the Southern sky including those we have yet to identify. For many purposes, the lack of a reliable initial object classification will restrict scientific output as many sources will still need spectroscopic verification.% and other follow-up observations, a process that SDSS has been tackling for over two and a half decades.

While the 4-metre Multi-Object Spectrograph Telescope \citep[4MOST,][]{de19} aims to obtain wide-area spectroscopic data for 13 million galaxies and AGN, it has yet to start and will only provide near-hemispheric coverage by 2030. Thus it is still relevant to explore hitherto underutilised datasets that might offer an opportunity to create a substantially complete sample of low-luminosity AGN.}

% There exist a spectro survey 6dFGS, and various attempts with different intentions and purpose to make an AGN cata
Preliminary work to solve this problem utilised the Six-degree Field Galaxy Survey \citep[6dFGS,][]{jones09}. This is a flux limited redshift survey of extended galaxies brighter than a K-band magnitude of $\sim13$, complemented by several smaller samples of special-purpose targets. In its final release, DR3, it had measured the redshifts of 110,256 extragalactic sources for the first time and delivered a catalogue of 125,071 galaxies (across 136,304 spectra) with a median redshift of 0.053. 

Various studies have classified and selected AGN in the 6dFGS sample for different scientific purposes. \cite{mahony10} visually inspected 1715 X-ray sources from the ROSAT All Sky Survey Bright Source Catalogue that were also observed by 6dFGS and had reliable redshifts, providing a set of homogeneous optical spectra. They identified 81\% of the sample to be AGN, 6\% as absorption-line galaxies, and 13\% as stars. Meanwhile, \cite{chen18} performed line-fitting on statistically selected spectra with strong H$\beta$ and [O{\sc iii}] $\lambda5007$ emission to identify Narrow-line Seyfert-1 (NLS1) galaxies. Finally, \cite{masci10} also constructed an AGN sample from the infra-red photometry of the Two Micron All-Sky Survey \citep[2MASS,][]{2MASS}. In all cases, the resulting lists were constructed for a specific science case and thus included only a subset of AGN within the 6dFGS sample.

% This work, we present the most exhaustive list within 6dFGS by constructing it mainly visually
An exhaustive AGN list is difficult to construct from 6dFGS, largely because the survey and instrument were designed for quantity over quality; spectra were taken for a redshift measurement and not for astrophysical characterisation. The plate scale of the wide-field UK Schmidt Telescope implied large fibre apertures of 6.7 arcsec diameter. The spectra also tend to have low signal-to-noise ratios, poor telluric subtraction, and are not flux calibrated. Thus, a simple automated process may struggle to select AGN from 6dFGS. However, these issues can be handled when done in conjunction with visual inspection. A combination of automated algorithms and visual verification was also used for the construction of the latest SDSS quasar catalogue \citep{lyke20}. 

% This will facilitate future development of AGN research in the southern sky
In this paper, we present {\refbf an extensive and reliable list} of broad emission line AGN (BEL AGN) within 6dFGS, verified through visual inspection and starting from a generously selected parent sample. This is the first stage of our effort to fully classify all optically-selected AGN from 6dFGS, with a planned second stage to identify all Type 2 AGN. {\refbf The main intention for this BEL AGN list is to identify low-luminosity AGN in the K-band flux-limited galaxy sample that forms the bulk of 6dFGS and reaches redshifts around 0.2. As a byproduct, we will also identify higher-redshift quasars among the available spectra, which were an incomplete list of auxiliary targets obtained as part of the broader 6dFGS observations.} The catalogue is intended to aid future research into Southern AGN and complementary surveys. 

The paper is structured as follows: Section 2 provides an overview of 6dFGS and Section 3 describes the spectrum handling processes. The catalogue creation is detailed in section 4 with the structure of our data product given in Section 5. In Section 6, the statistical properties are explored and compared to other catalogues. We summarise our work in Section 7. 

We adopt a flat $\Lambda$CDM cosmology with $\Omega_\Lambda=0.7$ and $H_0=70$~km~s$^{-1}$~Mpc$^{-1}$. Optical magnitudes are in the AB system, and IR magnitudes are in the Vega system.

\section{6dFGS overview}
From 2001 to 2006, the 6dFGS obtained 136,304 spectra using the UK Schmidt Telescope and the Six-degree Field multi-object fibre spectrograph. The survey targeted objects in a 17,000 deg$^2$ sky area in the Southern hemisphere at |b|>10 degrees (twice the sky area of SDSS DR7). 

\subsection{Survey Targets}
6dFGS is a compilation of sources from a range of science programs. The targeted sources in 6dFGS are selected using 2MASS, DENIS, and SuperCOSMOS magnitudes, spanning the K, H, J, I, r$_F$ and b$_J$ bands. The total sample contains sources that meet the magnitude limits of each band, and they are differentiated through the value of the {\sc prog\_id} column in the catalogue, as summarised in Table~\ref{tab:progid}.

\begin{table}
\centering
\caption{Table of the samples that make up the total list of 6dFGS targets and their associated {\sc prog\_id}.}
\label{tab:progid}
\resizebox{0.35\textwidth}{!}{%
\begin{tabular}{l|l}
\hline
{\sc prog\_id} & Sample \\
\hline 
\multicolumn{2}{c}{Flux Limited} \\
\hline 
1 & 2MASS Kext $\leq$ 13.00 \\
3 & 2MASS Hext $\leq$ 13.05 \\
4 & 2MASS Jext $\leq$ 13.75 \\
5 & DENIS J $\leq$ 14.00 \\
6 & DENIS I $\leq$ 14.85 \\
7 & SuperCOSMOS r$_F \leq$ 15.6 \\
8  & SuperCOSMOS b$_J \leq$ 16.75 \\
\hline 
\multicolumn{2}{c}{Auxiliary} \\
\hline 
78 & UKST \\
90 & Shapeley supercluster \\
109 & Horologium Sample \\
113 & ROSAT All-Sky Survey \\
116 & 2MASS Red AGN \\
119 & HIPASS \\
125 & SUMSS/NVSS radio \\
126 & IRAS FSC \\
129 & Hamburg-ESO QSOs \\
130 & NRAO-VLA QSOs \\
999 & Unassigned \\
\hline
\end{tabular}%
}
\end{table}

Kext, Hext, and Jext refers to the total magnitudes reported in the 2MASS extended source catalogue \citep{jarrett00}, which differ from those in the point source catalogue \citep{cutri03}. However, 6dFGS constructed a different set of magnitudes from 2MASS that accounts for the bias against low-surface-brightness galaxies \citep[for details, see][]{jones04}. This makes the survey limits on Kext, Hext, and Jext only an approximation to the true magnitude limits. In addition to the flux limited sample, 6dFGS observed several categories of auxiliary targets, which are also listed in Table~\ref{tab:progid}. For this work, we used targets from all samples. 

\subsection{6dF Redshift estimation}\label{sec:methods:sub:6dfgsreview}
6dFGS employed a method for redshift estimation that was inherited from 2dFGRS. It was designed to deliver accurate and high quality measurements for a majority of spectra. There are two modes of redshift estimation: using either absorption or emission lines \citep[for full details see Section 7 of][]{colless01}. 

Absorption redshifts were measured {\refbf by a cross-correlation process with spectral templates from five NGC galaxies with different morphological types, using the following steps:} (1) continuum subtraction using a sixth-order polynomial fit to the continuum and masked sky lines, (2) removal of strong emission lines identified by data points that were 5$\sigma$ above the mean, (3) rebinning to a logarithmic wavelength scale, (4) tapering the ends of the spectra to zero, (5) Fourier transformation, (6) applying an exponential filter to reduce the residual continuum at low wavenumber and noise at high wavenumber. The quality of the absorption redshift was given as a value between one and four, depending on the strength of the best correlation fit in relation to other matches. This estimation was suitable for most spectra because 6dFGS is dominated by extended sources that have a significant contribution from a stellar population. 

Emission redshifts were estimated by searching for a multi-line match from a pool of candidate emission lines in each spectrum. The procedure was as follows: (1) the continuum and sky lines were removed, (2) peaks reaching above 3.3 times the root-mean-squared from the mean were flagged as candidate emission lines, (3) the strength of candidate lines was evaluated by fitting a narrow Gaussian with full width at half maximum (FWHM) between 0.7 to 7 pixels, (4) for spectra with more than two line candidates, more than one potential redshift can be calculated and set as the emission redshift. For those with only one emission line, the candidate line was assumed to be either [O{\sc iii}]$\lambda$5007\AA\ or H$\alpha$. The quality of the emission redshift was the number of emission lines with the value of four being the maximum. This mode of estimation works for emission line galaxies regardless of whether the absorption lines are visible. However, because the Gaussian used in the detection is narrow, this method may not accurately detect the BELs of AGN, which tend to have a FWHM $>7$ pixels.

Every spectrum goes through both modes of reshift estimation, and the resulting estimation with the best quality was chosen as the final redshift, with a priority on the absorption redshift. This also determines the quality flag, {\sc Q}, of the spectrum listed in the catalogue, which inherits the quality of the redshift estimation. In the catalogue, a spectrum with {\sc Q}=1 or {\sc Q}=2 is considered unreliable, {\sc Q}=3 is reliable, {\sc Q}=4 is high-quality and {\sc Q}=6 is a Galactic source. If both automated modes failed, the redshift was determined by visual inspection. The final redshift was listed with a heliocentric correction.  

\subsection{Challenges with 6dFGS spectra}\label{sec:overview:sub:spectra} 
6dFGS spectra are observed in two wavelength regimes at different times, using a V ($\lambda\lambda3900$-5600\AA) and R ($\lambda\lambda5400$-7500\AA) grating that are then spliced together to form a V\&R frame prior to redshift determination. To reiterate, these spectra are obtained for the purpose of redshift determination and not astrophysical measurements. They are not flux calibrated, are low in signal-to-noise, and have poor telluric subtraction. In addition, it is common for 6dFGS spectra to further suffer from the following issues:
\begin{itemize}
    \item Inconsistent wavelength calibration between V and R spectra. The two arms may show relative offsets of up to 500 km/s in extreme cases. As a consequence of this, the removal of the $\lambda$5577\AA\ sky line is affected. The inconsistency acts to delocalise this sky line during splicing, shifting it several pixels thereby increasing the affected range. Since this sky line may interfere with the H$\beta$ and [O{\sc iii}]$\lambda\lambda$4959,5007 lines, an automated removal of this line often removes too much information.
    \item Inconsistent flux calibration between the V and R spectra. In most spectra, the V and R arms have significantly different scaling, with the R arm being scaled down. In some cases, we also observe a mismatch in spectral slope or continuum level at the splicing seam. 
    \item {\refbf Fringing and poor background subtraction that results in artefacts mimicking spectral features including broad humps. However, this phenomenon affects only about 1\% of the spectra}. 
\end{itemize}

Some groups of spectra are also affected by fibre cross-talk; {\refbf in about 1\% of spectra, this effect is strong enough to suggest an incorrect redshift}. This is a known issue with 6dFGS {\refbf discussed on the project website\footnote{\url{http://www-wfau.roe.ac.uk/6dFGS/xtalk/}}, where an incomplete list of affected spectra is included}. Affected spectra display features from another spectrum, resulting in two sets of spectral lines and an ambiguous redshift. Typically, one set of lines in an affected spectrum is the exact same set as that in the spectrum with a neighbouring {\sc spec\_ID} (unique spectrum identifier in the 6dFGS database). 

\begin{figure*}
    \centering
    \includegraphics[width=0.9\textwidth]{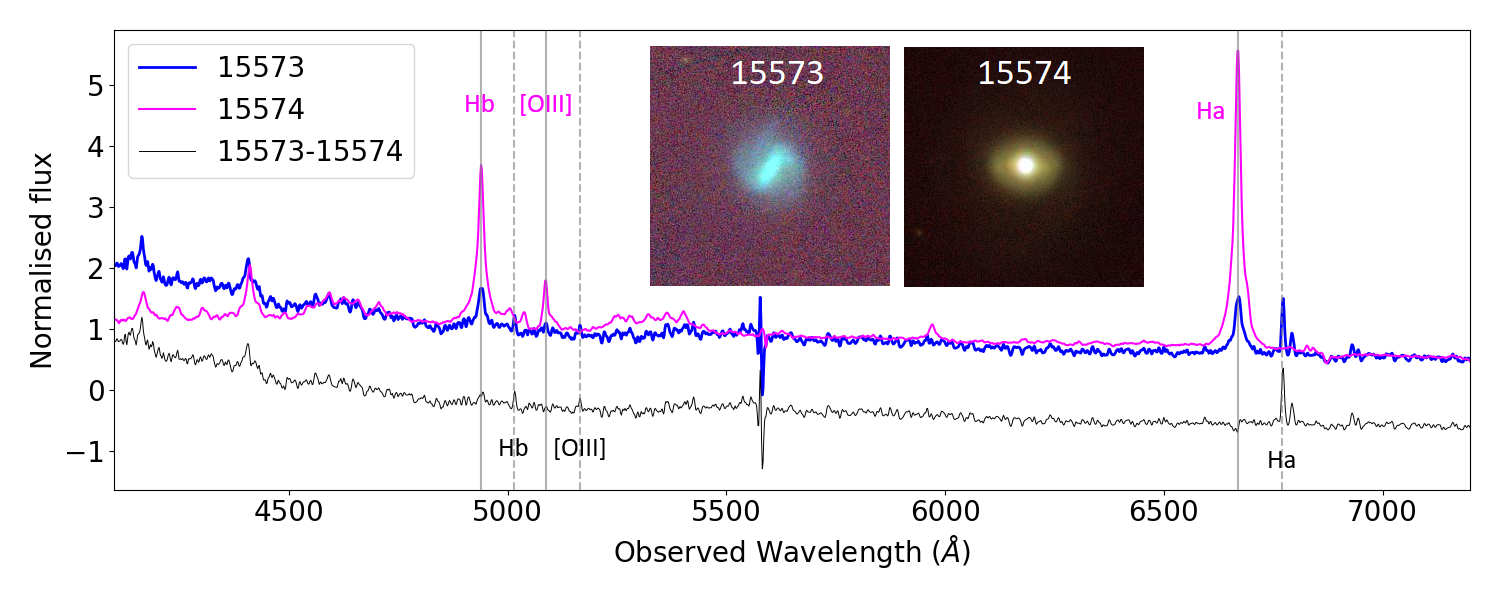}
    \caption{Figure showing an example of a fibre cross-talk. Blue thick line is the affected spectrum with {\sc spec\_id}=15573 and has emission lines indicated by vertical dashed grey line at $z=0.0317$. Magenta line is the spectrum with {\sc spec\_id}=15574 and has emission lines indicated by vertical solid line at $z=0.0160$. Emission lines at this redshift are also indicated with magenta font. The set of emission lines from 15574 is duplicated onto 15573. Black thin line is 15574 subtracted from 15573 with some arbitrary scaling to remove H$\beta$ and H$\alpha$, showing that the lines are duplicated, although not consistently scaled since H$\gamma$ is not cleanly removed. Two insert images are included. These are the PanSTARRS image cutout labelled accordingly.}
    \label{fig:chp:6dfagn:15573cross}
\end{figure*}

In Figure~\ref{fig:chp:6dfagn:15573cross}, we show an example involving the affected spectrum of {\sc spec\_ID}$=$15573 at $z=0.0317$ and its neighbour in the fibre sequence, 15574 at $z=0.016$. The latter is a spectrum of a typical NLS1 galaxy, but its emission lines are also found in 15573. The PanSTARRS \citep{flewelling20} image cutout reveals an isolated galaxy with no sign of contamination from the surrounding region. It is unlikely that the contamination comes from the same line-of-sight given the closeness of the redshifts of the two emission line systems. The 6dFGS position coordinates of these two spectra are 2.5 degrees apart, meaning that these two sources cannot have been within the same fibre, clearly demonstrating an issue of fibre cross-talk. This issue is over-represented in BEL AGN within the 6dFGS sample and complicates the redshift estimations of spectra. It is therefore crucial that these are carefully considered so non-AGN spectra are not included in our AGN count.

\section{Spectral line fitting}\label{pyqsofit}
The tool used to fit all spectra for this work was a modified version of the python package {\sc PyQSOFit}. The original version and technical details can be found in \url{https://github.com/legolason/PyQSOFit}, and it utilises the fitting algorithm {\sc kmpfit} from \url{https://www.astro.rug.nl/software/kapteyn/kmpfittutorial.html}. The modification added the use of skewed Gaussian and Voigt profiles, and also improved on the functionality to group fitted components such that they share emission line properties. The modified version is named {\sc PyQSOFit\_SBL} and can be found in \url{https://github.com/JackHon55/PyQSOFit_SBL}. In the following we describe the algorithm, and then its application to this work.

\subsection{Continuum fitting}\label{sec:pyqsofit:contfit}
{\sc PyQSOFit} starts by fitting the continuum. The code was designed to simultaneously estimate the galaxy and AGN continuum. The fitting is done with wavelength ranges containing galaxy and AGN emission features masked. These are the [O{\sc ii}]$\lambda3727$, [O{\sc iii}]$\lambda\lambda$4959,5007, [S{\sc ii}]$\lambda\lambda6716,6731$, the Balmer lines, the He{\sc i}$\lambda4686$, and He{\sc ii}$\lambda5876$ lines. During a fit, it considers:
\begin{itemize}
    \item The starlight from the host galaxy via template fitting and principal component analysis \citep{abdi10}.
    \item The AGN power-law continuum via a power-law component
    \item The AGN Fe{\sc ii} multiplet emissions via a template from \cite{boroson92} for emissions around H$\beta$ and from \cite{mejia16} for emissions around Mg{\sc ii}
    \item Any inconsistencies from templates via a 3rd order polynomial
\end{itemize}

Modelling the Fe emission with only one template will lead to inaccurate continuum fitting. While doing so will underestimate the error of the continuum fitting, most 6dFGS spectra simply do not have sufficient quality to benefit from a multi-template Fe emission modelling. The Fe emission is often spectrally unresolved, or the spectra have insufficient signal to noise.

\subsection{Emission line fitting}

{\refbf Low-redshift AGN show multiple narrow emission lines and potentially broad components in some of these lines. Here, the focus is on the set of lines crucial to BEL AGN typing. Our list of broad emission lines includes H$\alpha$ and H$\beta$ as well as Mg{\sc ii}$\lambda$2799. As the focus of this work is low-redshift objects we do not fit the CIV emission line that may appear in the spectra of a handful of high-redshift objects. The CIV line is also often asymmetric and more difficult to fit. 

The narrow emission lines (NELs) include H$\alpha$ and H$\beta$ as well as the [O{\sc iii}]$\lambda\lambda4959,5007$ doublet and the [N{\sc ii}]$\lambda\lambda6549,6585$ doublet. [O{\sc iii}] is required for AGN typing (see Section~\ref{SEC:belagnidentification}). Since this line often shows blueshifted wings, the FWHM is allowed to range up to 1,200 km s$^{-1}$. The rest of the NELs are blended with BELs, and have to be fitted to accurately model the impacted BELs. An upper limit of FWHM $<$ 700 km s$^{-1}$ limit is used for these narrow line fits.

Some of these lines are simultaneously fit in groups, always with joint constraints on line shapes (FWHM, velocity offset, and skew). In the case of doublets, the flux ratio is also fixed. All fitted emission lines are listed in Table~\ref{tab:mainlines} together with their constraints and groupings.

A skewed Gaussian was chosen for the broad Balmer lines, where the kurtosis measures the asymmetry of the profile. When a double-peaked profile was resolved, we used two sets of broad components, where one had a positive velocity offset and the other a negative one}. There were also cases where the profiles of the broad Balmer lines were different, either due to a spectral artefact, wavelength calibration error, or intrinsic variance. {\refbf In these cases, we refit these as individual lines instead of using joint constraints}. For any broad profiles that we were unable to deblend with a satisfactory model, such as those for narrow line Seyfert-1 galaxies, we used a skewed Voigt profile instead. 

{\refbf Finally, we model the entire Mg{\sc ii} emission-line doublet with a single skewed Voigt profile. Given the low spectral resolution and data quality of 6dF, we cannot deblend the components of the line, and intend to use the line information only to determine the redshift in the absence of any other lines. }

\begin{table*}
\centering
\caption{Table listing the groups of lines that are fitted in this work. 1st column is the names of the lines, 2nd is the spectral redshift at which the lines are considered, 3rd is the wavelengths of the group listed in order, 4th, 5th, and 6th describe the Gaussian model used in fitting, 7th column states the theoretical ratio that relates the two line fluxes within the group.}
\label{tab:mainlines}
\resizebox{0.95\textwidth}{!}{%
\begin{tabular}{ll|llllll}
\hline
Group & Lines & z range & Wavelength (\AA) & Type & Model & FWHM (km/s) & Flux Ratio \\ \hline
\ [O{\sc iii}] & \ [O{\sc iii}] & $<0.5$ & 4959, 5007 & Narrow & Symmetric & $<1200$ & 1/3 \\
\ [N{\sc ii}] & \ [N{\sc ii}]  & $<0.5$ & 6549, 6585 & Narrow & Symmetric & $<700$  & 1/3 \\
Balmer-NEL & H$\beta$, H$\alpha$ & $<0.5$ & 4861, 6563 & Narrow & Symmetric & $<700$ & - \\
Balmer-BEL & H$\beta$, H$\alpha$ & $<0.5$ & 4861, 6563 & Broad & Skewed & $>1200$ & - \\
- & Mg{\sc ii} & $0.5-1.5$ & 2799 & Broad & Skewed Voigt & $>1200$ & - \\
\hline
\end{tabular}%
}
\end{table*}

\subsection{Application to 6dFGS}\label{fitting_problems}
{\refbf Any automated fitting algorithm needs supervision and tweaks on a case-by-case basis when the data quality is low and spectral artefacts may mislead the fitting}. Localised artefacts such as a noise feature or the 5577 skyline are either masked or modelled. When there is a mismatching calibration between V and R arms, the continuum in each arm is fitted separately. While most of the spectra {\refbf can be fitted with the addition of a 3rd order polynomial, a different strategy is used for the more extreme cases: } Instead of modelling the continuum using the listed components {\refbf  (see Section~\ref{sec:pyqsofit:contfit})}, pixels of high statistical significance are removed iteratively to mask emission and absorption features until a good fit to the underlying continuum remains. 

\begin{figure}
    \centering
    \includegraphics[width=0.47\textwidth]{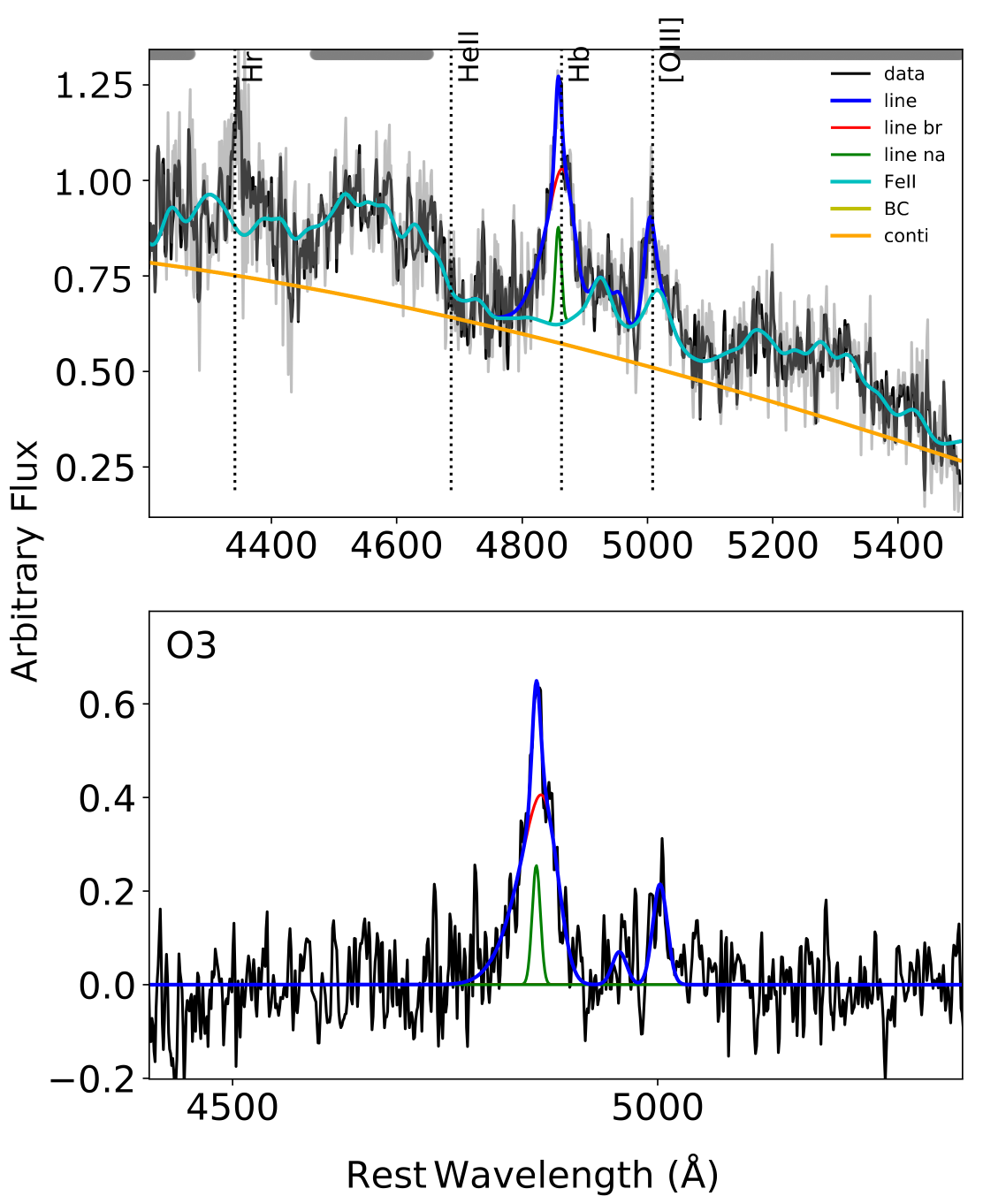}
    \caption{{\refbf A PyQSOFit visual output, showing an example of what is considered a good fit for a spectrum with signal-to-noise of 16.7. This is an average quality spectra as the median signal-to-noise is 16. PyQSOFit output shows two panels. \textit{Top:} highlights the continuum fitting with the yellow line. The wavelength windows used for fitting are shown as grey lines at the top. \textit{Bottom:} highlights the emission line fitting on a continuum subtracted spectrum. This fit was considered good as the subtracted continuum is consistent around 0 at the bottom panel, and the modelled emission line accurately traces the true emission line profile.}}
    \label{fig:chp:methods:visualex}
\end{figure}

\begin{figure*}
    \centering
    \includegraphics[width=0.75\textwidth]{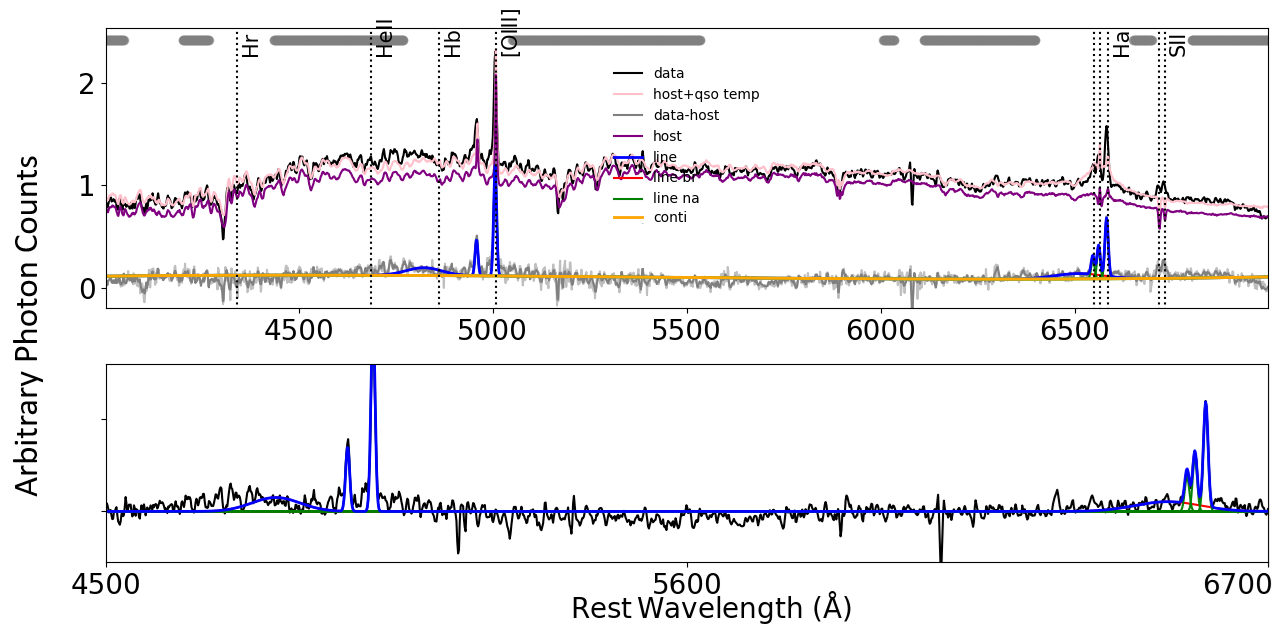}
    \includegraphics[width=0.75\textwidth]{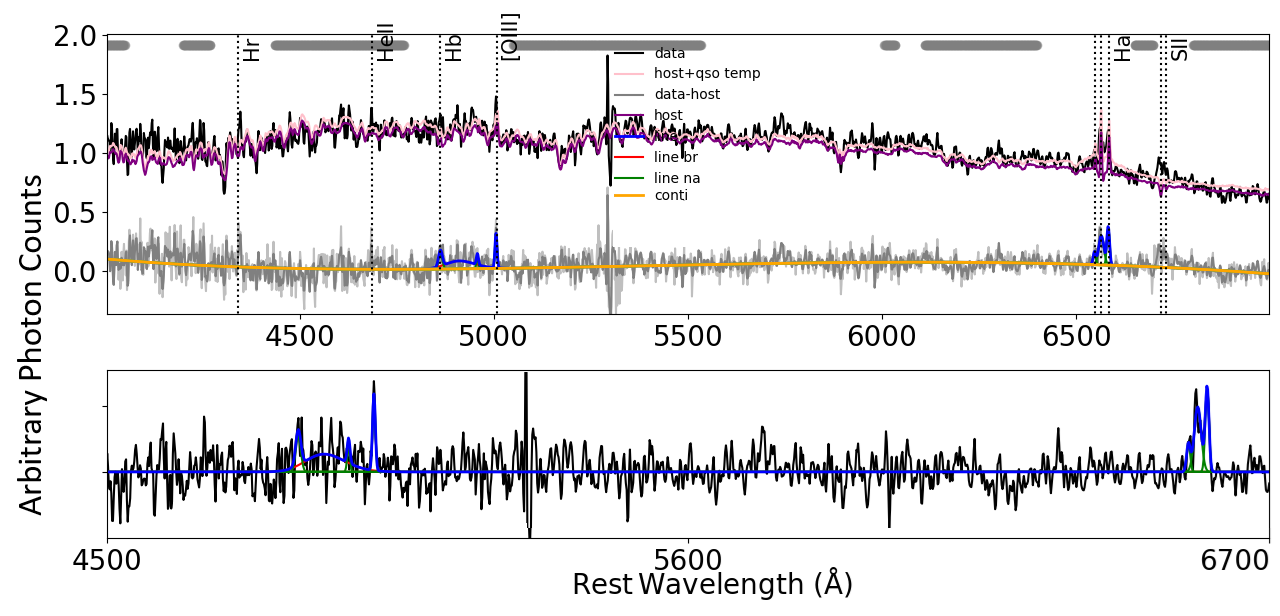}
    \caption{Figure showing examples of spurious broad lines that are detected. \textit{Top} is an example of spurious H$\alpha$ and H$\beta$, where the modelled line is fitting the residual caused by imperfect continuum subtraction to the left of H$\alpha$. \textit{Bottom:} is an example of a fit with a spurious broad H$\beta$ profile. {\refbf PyQSOFit considers the noise as a significant broad component. However, this spectrum is not an AGN because H$\alpha$ lacks a similar broad component.}}
    \label{fig:chp:methods:fakehb}
\end{figure*}

This continuum fitting method is similar to the online spectrum fitting tool known as Marz \citep{hinton16}. {\refbf Whereas Marz utilises polynomial fitting, we opted to use Gaussian smoothing instead to increase the flexibility of the tool. The sigma for the smoothing is between 25 and 150 depending on the scale of the artificial feature in the problematic continuum. A large sigma is typically used for a wide hump, while a smaller sigma is used for narrow fringes. The value of sigma is modified by eye until a satisfactory continuum fit is achieved.

At $z<0.5$, the restframe wavelength range of 4000-7000\AA\ contains the relevant emission features. This range generally also encapsulates sufficient continuum for an accurate fit (2200-3800\AA\ for $0.5<z<1.5$). Visually, a `good' fit is obtained when the modelled line closely traces the mean of the noise at every part of the spectrum and especially around the BEL. Figure~\ref{fig:chp:methods:visualex} is an example of a `good' fit, and Figure~\ref{fig:chp:methods:fakehb} shows the opposite.
}

\subsection{BEL AGN identification}\label{SEC:belagnidentification}
At low redshift, BEL AGN are typically Seyfert galaxies. These are commonly categorised into sub-types based on the convention introduced by \cite{osterbrock81}. We adopt the convention from \cite{winkler92}, separating AGN into type 1.0$\rightarrow$1.2$\rightarrow$1.5$\rightarrow$1.8$\rightarrow$1.9$\rightarrow$2.0. The definitions of these types are defined as:
\begin{itemize}
    \item Type-1.0 \ \ \ \ \ \ \ \  5.0 $\leq$ R
    \item Type-1.2 \ \ \ \ \ \ \ \  2.0 $\leq$ R $<$ 5.0
    \item Type-1.5 \ \ \ \ \ \ \ \  0.333  $\leq$ R $<$ 2.0
    \item Type-1.8 \ \ \ \ \ \ \ \  R $<$ 0.333
    \item Type-1.9 \ \ \ \ \ \ \ \  no H$\beta$ BEL, only H$\alpha$
    \item Type-2.0 \ \ \ \ \ \ \ \  no H$\beta$ or H$\alpha$ BEL
\end{itemize}
where R is the ratio of H$\beta$ flux and [O{\sc iii}]5007$\lambda$ NEL flux. 
\begin{align}
    \mathrm{R} &= \dfrac{f(\mathrm{H}\beta_{\mathrm{BEL}})}{f([\mathrm{OIII}]5007\lambda)} 
\end{align}
However, in contrast to \cite{winkler92}, we use only the broad-line emission in the H$\beta$ line for more consistent comparison across all spectra which have  varying contributions from the host galaxy.

The necessity of the broad H$\alpha$ component is tested for these cases. By fitting only the H$\alpha$ region between the rest wavelength 5500-7000\AA\ with and without the broad Gaussian, the broad H$\alpha$ is considered to truly exist if the following were satisfied:

\begin{itemize}
    \item {\refbf At close to zero flux, the width} of the broad Gaussian is wider than the three narrow lines, [N{\sc ii}]$\lambda\lambda6549,6585$, and H$\alpha$
    \item {\refbf The fitted component models a visually obvious broad profile. Specifically, the area encapsulated by this component is not dominated by noise or caused by imperfect continuum subtraction (see e.g., Figure~\ref{fig:chp:methods:fakehb}). These cases are ruled out by checking the {\sc PyQSOFit} continuum fitting against our continuum fitting tool.}
    \item The removal of the broad Gaussian significantly impacts the fitting around the [N{\sc ii}]$\lambda6549$ line, i.e. the removal should result in a residual that exceeds the noise level.
    \item {\refbf The broad profile cannot be explained by broad [N{\sc ii}]. This is only applicable if there are broad [O{\sc iii}] as well. To determine this factor, broad [N{\sc ii}] components are fitted in place of a broad H$\alpha$. We rule out a BEL AGN if the broad [N{\sc ii}] width is comparable to the broad [O{\sc iii}].}
\end{itemize}

This list is only a general description for our test. Since the quality of spectra is low, the determination of BEL AGN will always be on a case by case basis. A common factor that hinders the determination of broad H$\alpha$ is the B-band telluric absorption around $\lambda$6885. This contaminates the H$\alpha$ profile at low redshifts, either by hiding the broad component or by tricking the continuum fitting into an incorrect solution that results in a broad component after subtraction. These cases were still considered but flagged as possibly erroneous.

Many BEL AGN will not have H$\alpha$ within the 6dFGS wavelength range, either due to their redshift or the R arm being unavailable. Determination of Type-1.9s is therefore not possible, and only Type-1.0s to Type-1.8s can be identified. For these, determining the existence of H$\beta$ becomes important. Bright Type-1.0s and 1.2s are easily recognisable much like the strong H$\alpha$ BEL. For the weaker sources, the host galaxy contribution and the Fe{\sc ii} emission become important to model accurately as the profile of the broad H$\beta$ line heavily depends on those contributions. The criteria for determining a true H$\beta$ line are the same as H$\alpha$, with an emphasis on ruling out broad profiles formed by imperfect continuum subtraction. These usually appear as good fits with a large FWHM, but vanish when the continuum is modelled with our continuum fitting tool (see Figure~\ref{fig:chp:methods:fakehb} e.g.). Finally, determination of Mg{\sc ii} BEL for sources at $z>0.5$ closely follows that for H$\beta$.

There are cases where a spectra visually has a broad line emission, but a good fit is not possible. For example, this can be due to a partial interruption by the $\lambda5577$\AA\  skyline or B-band telluric, where one wing of the broad line is visible but the other is not recoverable. These cases will be labelled with `bad' in the catalogue.

\section{Catalogue Construction}\label{sec:methods}
{\refbf
The catalogue of BEL AGN is constructed from a parent sample of 117,610 6dFGS spectra that have a quality flag {\sc Q}=3 or 4. For ease of discussion, we define "sky-completeness" as the fraction of AGN in the catalogue compared to that in the actual sky. This completeness follows the same statistics as that of the galaxies within the parent sample. This will be discussed in Section~\ref{sec:discussion:sub:completeness}.

For this section, we focus on the "catalogue-completeness", which is the fraction of AGN in the catalogue compared to that in the parent sample. The aim is for all BEL AGN to be selected with a catalogue completion rate that is as close to 100\% as possible, with few to no contaminants. This means that we have to consider all the spectra with the problems described in Section~\ref{sec:overview:sub:spectra} rather than deselecting them, as well as visually verifying each selected BEL spectrum. 
}

The procedure for identifying the BEL AGN proceeds in two steps: firstly, all emission line galaxies were identified and their redshifts verified; secondly, those sources with broad emission lines were selected and in each case visually confirmed. The reliability of our selection is discussed in Sec \ref{sec:discussion:sub:reliability}.

In the following sections, we describe our method in detail. The .fits file of each spectrum was obtained through the publicly accessible domain \url{http://www-wfau.roe.ac.uk/6dFGS/}.

\subsection{Emission line galaxy identification - $z<0.5$}\label{sec:method:sub:RFfinder}
\begin{figure}
    \centering
    \includegraphics[width=0.47\textwidth]{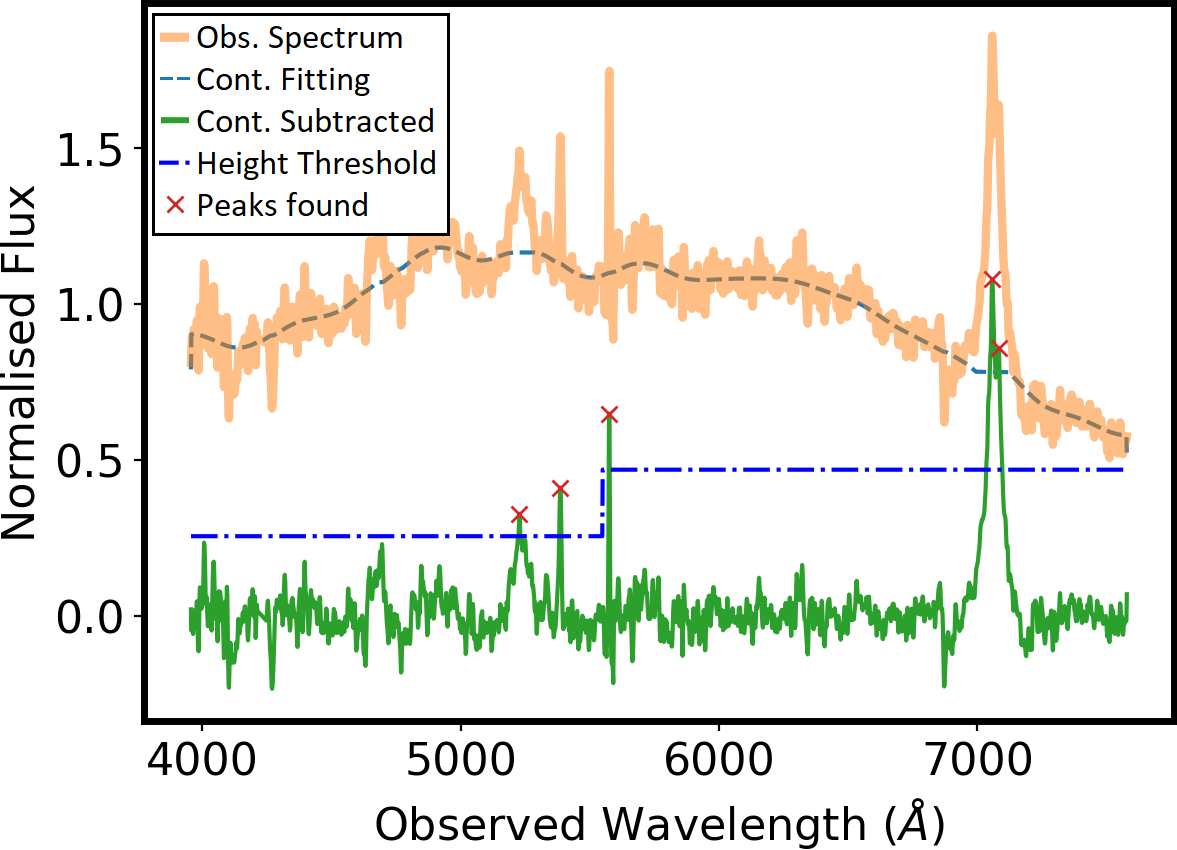}
    \caption{Figure presenting an example from the emission redshift estimation algorithm. The plot shows the continuum subtracted result, the split in detection threshold in the V and R parts of the spectrum, and the candidate lines found. One of the candidate line is the 5577\AA\ sky line, while the others found will be identified as H$\beta$, [O{\sc ii}]5007\AA, H$\alpha$ and [N{\sc ii}]6585\AA}
    \label{fig:rfex}
\end{figure}

{\refbf
An accurate method to select emission line galaxies from the parent sample would be to utilise the already existing redshift estimate of 6dFGS. To do so, emission redshifts of each spectrum are re-estimated then compared to the reported redshift of 6dFGS. Any spectrum that matches within a 10\% margin will be considered an emission line galaxy candidate. It is expected that the margin of error grows with redshift as the accuracy of redshift estimation would be decreasing. This is because broad lines begin to dominate the spectrum. The entire procedure is described in detail in the following subsections. An example of an AGN selected by our algorithm is shown in Figure~\ref{fig:rfex}.
}

\subsubsection{Data cleaning}
The spectra are homogenised with negative and unbounded pixels removed, then normalised to the mean. The continuum is also subtracted. Rather than fitting with a multiple component model, the self-written continuum fitting tool (see end of Section~\ref{fitting_problems}) was used in order to quickly process the large number of spectra. There are two caveats in using this tool. 

Firstly, the continuum around broad lines is not perfectly accounted for and there will be some amount of area loss to the broad lines. This lowers the peak value and makes it harder to detect spectra with a weak H$\beta$ BEL as the sole feature. These cases can be recovered and will also be discussed in Section~\ref{sec:methods:sub:wrong}. 

Secondly, the host galaxy contribution was ignored. This is not unreasonable as emission lines listed in Table~\ref{tab:mainlines} are the important lines for emission line galaxy recognition. These lines are all apparent without subtracting the galaxy. The exception to this is the narrow H$\beta$, but recovering this line is difficult and often unnecessary for the purpose of this procedure.

Finally, the $\lambda5577$ skyline is not masked. This is to preserve spectra where the interference from the skyline makes them unrecognisable as emission line galaxies after masking.

\subsubsection{Detecting emission features}
The {\sc find\_peaks} function from the {\sc scipy} package is used to detect emission features. The noise levels in the V and R arms of 6dFGS spectra are often dissimilar. Therefore, these two arms are processed separately. Rather than defining a threshold at which features are considered statistically significant based on variance, a quota of detected features was set. This was done with the intention of maximising completeness and with a way of discarding a large number of contaminants. Starting with a 3$\sigma$ threshold, we increase or decrease this threshold until both the V and R arms have between three to twenty detected features. The lower limit is a necessary requirement for estimating redshift via emission, while the upper limit has been empirically tested, where increasing it further resulted in longer computational times with no significant return. 

The minimum distance between features was also set to 27 pixels or $\sim44$\AA. This distance ensures that {\sc find\_peaks} does not fill up the quota by densely selecting features in a noisy region. This limit does not affect the emission lines in Table~\ref{tab:mainlines}. 

\subsubsection{Emission Redshift Estimation and Comparing to 6dFGS}
{\refbf
Within the set of three to twenty detected features, a ``guess redshift'' can be calculated for each pair. This is based on the fact that the ratio of the wavelength between two emission line remains the same irrespective of the spectrum's redshift. 
}

Based on the description from Section~\ref{sec:methods:sub:6dfgsreview}, 6dFGS provides (1) the absorption redshift for a spectrum with significant absorption and emission lines, (2) the emission redshift for a spectrum with only significantly narrow emission lines, and (3) either emission redshift or manual redshift for spectrum with only BEL. If any ``guess redshifts'' matches with 6dFGS, the spectrum is almost certainly an emission line galaxy. However, a mismatch in case (3) could indicate a misidentification by the 6dFGS team. These redshift error cases, along with those where the redshifts don't match, as well as issues due to the 5577\AA\ skyline, will all be discussed in Section~\ref{sec:methods:sub:wrong}. 

About $25\%$ of spectra do not fullfil the three to twenty range quota. In some of these, it is because they only have a single obvious emission feature. These can often be an AGN at $z\sim0.3$ where the only emission line in the 6dFGS spectral range is H$\beta$. They can also be spectra with large artefacts that cause all remaining emission features to appear weak. To recover them, we test if an emission feature exists (above 3$\sigma$) in a 40 pixel or $\sim66$\AA\  window around H$\beta$, [O{\sc iii}]$\lambda$5007, and H$\alpha$. If an emission feature is present, then these spectra will not be discarded.

The number of spectra selected as emission line galaxies through this procedure is 66,848 out of the 136,304 spectra. This corresponds to 49\% of our original sample. 

\subsection{Broad line AGN Classification - $z<0.5$}\label{sec:methods:sub:lowzbel}
To select for BEL AGN, the emission line galaxies are fitted with {\sc PyQSOFit}. However, acquiring a good fit for the full 66,848 spectra is labour-intensive due to the previously discussed complications with 6dFGS spectra. This makes the task impractical and also unfruitful, considering that only $\sim20\%$ of bright galaxies are AGN \citep{barthel06}, and only a smaller fraction of those have broad lines. Furthermore, this sample is also contaminated by non-emission sources. 

Therefore, we have to begin by removing the obvious contaminants as well as pruning the sample down to a workable size. {\refbf Often, a draft fit to a broad line spectrum does not stray far from the good fit. This is because the probability that one of the 6dFGS issues lies on top of H$\beta$ or H$\alpha$ is low.} Therefore, the line fitting results from a draft fit can be used to inform which spectra to discard by creating selection cuts.

These cuts were constructed using a sample of common sources within the Million Quasar catalogue v7.4 \citep[MILLIQUAS,][]{flesch21}. For this purpose, the catalogue version used is not significant. This work started in the year 2021 and has been using version 7.4 but with references from `2MAGN' and `6dAGN' removed. AGN with these tags refer to the work from \cite{zaw19} and \cite {chen22}, which are other 6dFGS classification attempts which are completely automated. We have noted issues with their approach and will compare our results to theirs in Section~\ref{sec:discussion:sub:2magn}. 

There are 1,575 spectra within 10'' arcsec of the MILLIQUAS AGN at $z\leq0.5$. We fitted all of these spectra and were able to identify 677 BEL AGN with a good fit from PyQSOFit. In addition, there are also 110 sources where we can identify the presence of a BEL, but had issues that interfered with acquiring reliable line measurements to some of the fitted lines.

From those with good fits, three categories of broad line spectra were found and they can be defined with three values, R = H$\beta$(broad)/[O{\sc iii}], the same parameter for AGN typing, A = H$\alpha$(broad)/[O{\sc iii}] which is the analogue of R for H$\alpha$, and Y = H$\alpha$(broad)/H$\alpha$(narrow).

\begin{itemize}
    \item Type-1.0 to 1.8 with H$\alpha$ - These were spectra where R was significant and H$\alpha$ was within the spectral range. This subset was defined as R$\geq0.1$ with broad H$\beta$ and H$\alpha$ fluxes being non-zero. 
    \item Type-1.0 to 1.8 without H$\alpha$ - These were spectra where R was significant but H$\alpha$ was out of the spectral range. This subset was defined as R$\geq0.1$ with broad H$\beta$ flux being non-zero, but zero flux for H$\alpha$
    \item Type-1.9 sources - These were spectra with no broad H$\beta$ and therefore a very small R. This subset was defined as A$\geq0.1$, Y$\geq0.5$ and a non-zero flux for broad H$\alpha$. 
\end{itemize}

We then place the sample of 66,848 into each of these categories and identified 740 BEL AGN at $z<0.1$, 934 at $0.1<z<0.5$, and 163 'bad' cases within $z<0.5$, for a total of 1837 unique AGN sources. Those that did not belong in these categories will be discussed in Section~\ref{sec:methods:sub:wrong}.

\subsection{Broad line AGN Classification - $z\geq0.5$}\label{sec:methods:sub:belagnhigh}
At $z\geq0.5$, the 6dFGS is dominated by the targets from the auxiliary sample and no longer restricted to extended galaxies. Stellar emission for this subset of spectra will therefore be insignificant because of the higher redshift combined with the inclusion of point sources. The 6dFGS absorption redshift becomes irrelevant. In addition, the emission lines observable by 6dFGS at this redshift tend to be the broad quasar lines such as Mg{\sc ii}, C{\sc iv}, and Ly$\alpha$. Therefore, the 6dFGS emission redshift also becomes irrelevant because their method is only sensitive to narrow lines.  Spectra in this subset have a good quality 6dFGS redshift as the 6dFGS team had to visually inspect them. This means that most 6dFGS spectra at $z>0.5$ are highly likely to be AGN.

Fitting all spectra in this subset and acquiring good fits, 454 AGN at $0.5<z<1.5$ was found with a Mg{\sc ii} line. There are also 17 `bad' spectra, and 207 high redshift QSO between $z=1.5$ and $z=3.793$ that has been noted through visual inspection. This is total of 678 AGN at $z>0.5$. {\refbf These spectra are included in the catalogue to achieve a high catalogue-completeness. Their sky-completeness is complicated as it depends on the parameters of their respective surveys. As such, it is not recommended to use these as a high sky-complete sample.}

\subsection{AGN with incorrect 6dFGS redshifts}\label{sec:methods:sub:wrong}
A misidentification of redshift can occur to any spectra. To identify all of them would require a visual inspection of all of the 6dFGS sample. Looking at the 1,575 MILLIQUAS common sources at $z<0.5$, there are 18 emission line sources with a mismatch of redshifts reported by 6dFGS and MILLIQUAS. The magnitude of these differences ranges from 0.015 to 3.597 and affects sources from an actual redshift of 0.053 to 3.8. 

In most of these cases, the redshifts were underestimated due to misidentifying a quasar broad emission line (i.e. Ly$\alpha$, C{\sc iii}, Mg{\sc ii}) as the [O{\sc ii}]$\lambda$3727 doublet. These occur due to several factors. Firstly, the 6dFGS emission redshift estimation does not account for these high redshift quasar emission lines. Secondly, the estimation should have failed and flagged the spectra for a manual redshift where the 6dFGS team would then correctly identify them. However, that must not have happened, because the profile of these broad lines can be narrow enough to satisfy the $0.7>$FWHM$>7$ pixel at 3$\sigma$ criterion that was required for 6dFGS emission redshift estimation.

The only case of overestimated redshift was where [O{\sc iii}]$\lambda$5007 was misidentified as the [O{\sc ii}]$\lambda$3727 doublet. This specific case was caused by a non-physical continuum shape that `hid' the [O{\sc iii}]$\lambda$4959, which would otherwise have resolved the degeneracy. 

Both of these types of cases involve the [O{\sc ii}]$\lambda$3727 doublet, therefore spectra with this line in the observable range were targeted for visual inspection. Specifically, we looked over the spectra after cleaning and without fitting for an indication of incorrect redshifts. For underestimation, this is presented as a broad line at the observed wavelength of [O{\sc ii}]$\lambda$3727. For overestimation, this is a combination of the [O{\sc iii}]$\lambda$5007 misidentified as the [O{\sc ii}]$\lambda$3727 doublet, H$\beta$ line that is not identified, and a H$\alpha$ line that is mistaken as H$\beta$. 

We define two subsets to visually inspect: 
\begin{enumerate}
    \item Spectra with $z>0.2$ that were rejected as emission line galaxies. These potentially have underestimated redshifts. A careful search of this subset also finds spectra with extremely weak or noisy H$\beta$ BEL and with no other narrow lines. These are spectra that may have been missed due to the continuum subtraction tool used during the emission line galaxy selection.
    \item Spectra with $z>0.4$ that were rejected as broad line spectra in the category of Type-1.0 to 1.8 without H$\alpha$. We previously rejected these because a `good fit' could be found and there was no clear indication that a broad line existed. These could have overestimated redshifts. This means that a good fit was not possible because we were fitting the H$\alpha$ region as H$\beta$, and noise has masked the line identification. By looking at the whole spectra before blueshifting, the nature of the spectrum can be verified with the presence of the Mg{\sc ii} line.
\end{enumerate}

{\refbf In addition, serendipitously detected BEL spectra with incorrect redshifts were also noted}. Most of these occur when classifying the $z>0.5$ BEL AGNs. Smaller redshift discrepancies ($|z|<0.1$) were not {\refbf corrected} as these were problems with spectral resolution or wavelength calibration rather than a misidentification of emission lines. 

In total, we found 38 BEL spectra with an incorrect 6dFGS redshift. The range of the 6dFGS redshift was from $z=0.00683$-1.54425, while the true range is from $z=0.053$-3.80. Among them, 9 are BEL AGN with $z<0.5$, 10 are Mg{\sc ii} spectra, 13 are high-z QSO with $z>0.5$, and 6 are labelled as 'bad'.

\subsection{More on 6dFGS fibre cross-talk}
While visually inspecting for good fits, we identified 11 cross-talk cases (listed at the top of Table~\ref{tab:chp:6dfagn:xtalklist}). A couple of them have BEL and the cross-talk only introduced additional narrow lines. These cases are included into the BEL AGN catalogue and are explicitly labelled. Since the effect can arbitrarily alter the continuum profiles, all measurements provided are likely to be inaccurate for these cases.

We also found cross-talk like issues that affect more than a pair of spectra. It is more accurate to describe these as plate contamination since these extend to a range of spectrum ids with fibres belonging to the same plate. There are three main types of plate contamination: 

\begin{enumerate}
    \item A top hat profile at $\sim7000$\AA\ rest wavelength. This fakes the appearance of a H$\alpha$ BEL and effectively masks any H$\alpha$ BEL. The broad line profile in such cases are not recoverable. 
    \item A bump at $\sim5000$\AA\ rest wavelength. It is not clear whether this is caused by incorrect raw data processing or is a cross-talk issue duplicating a BEL from one spectrum across multiple spectra. Affected spectra appear to have an unrealistically strong BEL, typically between the expected wavelength of H$\beta$ and H$\gamma$. 
    \item A set of narrow emission lines at a redshift of 0. These are interstellar medium contamination, with most spectra being affected by the Vela supernova remnant based on their close proximity. While 6dFGS avoids most of the Galactic plane with |b|>10 degrees, the Vela supernova remnant reaches to almost -15 degrees. The additional lines all have a redshift of 0, making this case easily identifiable and handled with little disruption to the emission lines.
\end{enumerate}

The ranges of affected {\sc spec\_ID} that we have identified are listed below in Table~\ref{tab:chp:6dfagn:xtalklist}. similar to the cross-talk cases, those with broad emission lines will be included into the catalogue, but their measurements will be inaccurate.

\begin{table*}
\centering
\caption{Table listing cross-talk cases that we have identified in this work. Pair cases are listed at the top, extended range cases are listed at the bottom. Those that are still included in the BEL AGN catalogue are commented with `Included'. Since this list only contains spectra that our algorithm picked up as potential BEL AGN, it is not an extensive list of all the affected cases.}
\label{tab:chp:6dfagn:xtalklist}
\resizebox{0.67\textwidth}{!}{%
\begin{tabular}{l|ccl}
\hline
Name & spec\_ID & Cross-talk spectra & Comments \\
\hline
g0101488-163843	& 7519 & 7520 & \\
g0239545-082404 & 15573 & 15574 & \\
g0308226-284821 & 17296 & 17295 & Subtle bump at H$\alpha$ \\
g0350192-221721 & 21459 & 21458 & Included \\
g0414221-082044 & 24756 & 24755 & \\
g1042221-191457	& 50168 & 50169 & \\
g1134475-361011	& 55639 & 55640 & \\
g1138589-380042 & 55641 & 55640 & \\
g1143318-182804 & 127973 & 127974 & \\
g1148026-184952 & 127975 & 127974 & \\
g1308296-324420 & 65008 & 65009 & Included \\
g1405106-310335	& 72453 & 72424 & \\
g1603175-021246 & 80670 & 80671 & Subtle bump at H$\alpha$ \\
g1604362-372238	& 134727 & 134728 & \\
g1636268-211835	& 82764 & 82765 & \\
g2107466-163707 & 99154 & 99155 & Included \\
g2203170-360649 & 105573 & 105574 & Included \\
g2247320-400349	& 110022 & ? & Included \\
g2335290-355515	& 139634 & 139635 & \\
\hline
Top hat profile at $\sim7000\AA$ & \multicolumn{2}{c}{45601-45621} & 45608 Included\\
Top hat profile at $\sim7000\AA$ & \multicolumn{2}{c}{115797-115819} & \\
Top hat profile at $\sim7000\AA$ & \multicolumn{2}{c}{115822-115870} & \\
Bump at $\sim5000\AA$ & \multicolumn{2}{c}{97176-97230} & \\
Bump at $\sim5000\AA$ & \multicolumn{2}{c}{110639-110645} & 110644, 110645 Included\\
Bump at $\sim5000\AA$ & \multicolumn{2}{c}{120792-120820} & \\
Vela supernova remnant & \multicolumn{2}{c}{31248-31297} & 31269 Included \\
Vela supernova remnant & \multicolumn{2}{c}{34384-34505} & \\
Vela supernova remnant & \multicolumn{2}{c}{37947-37948} & \\
Vela supernova remnant & \multicolumn{2}{c}{40268-40309} & \\
Vela supernova remnant & \multicolumn{2}{c}{40332-40423} & 40402 Included \\
Vela supernova remnant & \multicolumn{2}{c}{42986-43040} & 43020 Included \\
Vela supernova remnant & \multicolumn{2}{c}{43223-43264} & 43231 Included \\
Vela supernova remnant & \multicolumn{2}{c}{52571-52586} & \\
Vela supernova remnant & \multicolumn{2}{c}{82150-82211} & 82211 Included \\
Vela supernova remnant & \multicolumn{2}{c}{119294-119336} & 119295 Included \\
Vela supernova remnant & \multicolumn{2}{c}{126068-126071} & 126069 Included \\
Vela supernova remnant & \multicolumn{2}{c}{129787-129830} & 129787 Included \\ 
Vela supernova remnant & \multicolumn{2}{c}{137597-?} & 137597 Included \\
\hline
\end{tabular}%
}
\end{table*}

\section{Data product}
\begin{figure*}
    \centering
    \includegraphics[width=0.9\textwidth]{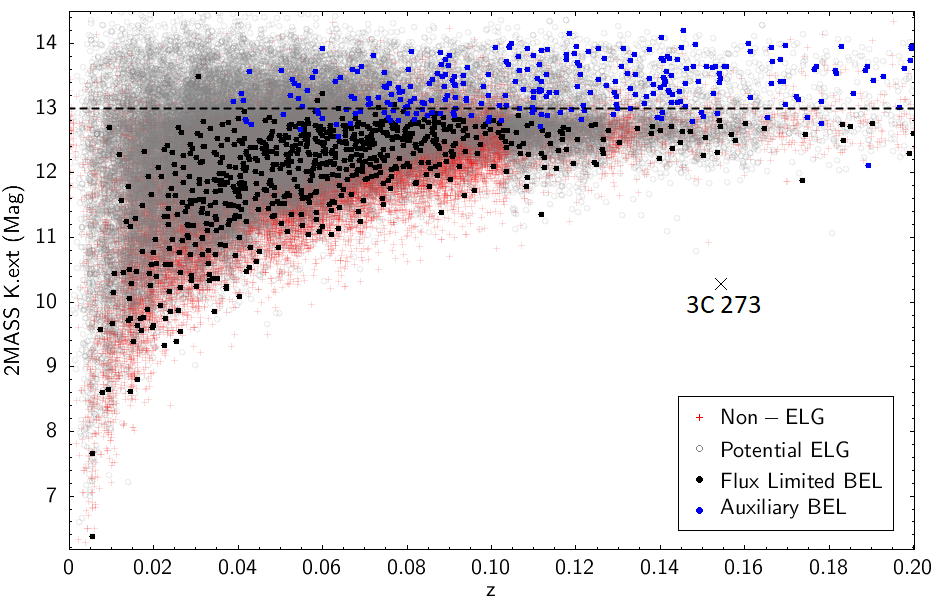}
    \caption{Figure presenting the distribution of our BEL AGN sample within $z<0.2$. We also indicate where 3C 273 lie relative to the 6dFGS sample.}
    \label{6dfbel:distribution}
\end{figure*}

The final catalogue has 2515 unique 6dFGS sources from a redshift of 0.0055 to 3.8, with a median of 0.207. Figure~\ref{6dfbel:distribution} shows the distribution of the flux limited BEL AGN sample compared to the 6dFGS sample within $z<0.2$, which is the bulk of the population. We list the columns names, units and simple descriptions in Table~\ref{tab:chp:6dfagn:colname}. Columns with longer descriptions are elaborated in the sections below. 

\begin{table*}
\centering
\caption{Table listing the columns in the final catalogue. * see description in text}
\label{tab:chp:6dfagn:colname}
\begin{tabular}{ll|l}
\hline
      Column Name   &  Column Unit & Description\\
\hline
      6dFGS\_name   & string & Unique 6dFGS name in format gHHMMSSS$\pm$DDMMSS\\
      RA            & float, Degrees & Right Ascension \\
      DEC           & float, Degrees & Declination \\
      prog\_id      & integer & 6dFGS identifier, see Table~\ref{tab:progid} \\
      spec\_id      & integer & 6dFGS identifier for each spectrum \\
      duplicates    & string & spec\_id's for a source with multiple spectra, separated by comma\\
      z             & float & Source Redshift\\
      z\_original   & float & The value of incorrect 6dFGS redshift if non-empty\\
      type          & string & * \\
      ext\_type     & string & * Name of external catalogue\\
      SNR           & float & * Signal-to-Noise estimate \\
      ebmv          & float & E(B-V) by Planck 2018 \citep{akrami20}\\
      Comment       & string & Various comments\\
      SMSS\_id      & integer & SkyMapper unique identifier\\
      dr4\_dist     & float, arcsec & Separation to closest SMSS dr4 match \\
      AGN\_to\_STAR & float & Estimated ratio of AGN to close neighbouring star (Section~\ref{SMSS:matching}) \\
      gaia\_obj     & integer & Number of objects within 4'' detected by Gaia \\
      gaia\_star    & integer & Number of stars within 4'' detected by Gaia \\
      mag67\_g       & float, AB mag & SkyMapper photometry constructed $g$-band mag (Section~\ref{sec:data:sub:lumest}) \\
      e\_mag67\_g    & float, AB mag & Error for above \\
      mag67\_r       & float, AB mag & SkyMapper photometry constructed $r$-band mag (Section~\ref{sec:data:sub:lumest})\\
      e\_mag67\_r    & float, AB mag & Error for above \\
      flux67\_g      & float, erg/s/cm/cm/\AA & SkyMapper photometry constructed $g$-band flux \\
      flux67\_r      & float, erg/s/cm/cm/\AA & SkyMapper photometry constructed $r$-band flux \\
      flux\_HbB\_est   & float, erg/s/cm/cm & Photometry estimated H$\beta$ BEL integrated flux \\
      flux\_O3\_est    & float, erg/s/cm/cm & Photometry estimated [O{\sc iii}] total integrated flux \\
      e\_flux\_HbB\_est  & float, erg/s/cm/cm & Corresponding error \\ 
      e\_flux\_O3\_est  & float, erg/s/cm/cm & Corresponding error \\     
      flux0\_HbB\_est   & float, erg/s/cm/cm & Corresponding de-reddened version \\
      flux0\_O3\_est    & float, erg/s/cm/cm & Corresponding de-reddened version \\
      logL\_HbB      & float, erg/s & log10 H$\beta$ BEL luminosity derived from de-redden estimated flux \\
      logL\_O3       & float, erg/s & log10 [O{\sc iii}]$\lambda$5007 luminosity derived from de-redden estimated flux \\
      e\_logL\_HbB      & float, erg/s & Corresponding error \\ 
      e\_logL\_O3       & float, erg/s & Corresponding error \\ 
      
      flux\_Mg\_est    & float, erg/s/cm/cm & Photometry estimated Mg{\sc ii} BEL integrated flux \\  
      e\_flux\_Mg\_est  & float, erg/s/cm/cm & Corresponding error \\ 
      flux0\_Mg\_est   & float, erg/s/cm/cm & Corresponding de-reddened version \\
      logL\_Mg      & float, erg/s & log10 Mg{\sc ii} BEL luminosity derived from de-redden estimated flux \\
      e\_logL\_Mg      & float, erg/s & Corresponding error \\ 
    
      r\_HbB\_O3    & float & flux0\_HbB\_est / flux0\_O3\_est \\ 
      
      FWHM\_X        & float, km/s & * Full width at half maximum for measured line \\
      e\_FWHM\_X     & float, km/s & Corresponding error \\
      skew\_X       & float & * Kurtosis value for measured line \\
      e\_skew\_X     & float & Corresponding error \\
      peak\_X     & float, \AA & * Peak value for measured line \\
      e\_peak\_X     & float, \AA & Corresponding error \\
      EW\_X       & float & * Equivalent Width of measured line \\
      e\_EW\_X     & float & Corresponding error \\
      
      2MASS\_id & string & Unique 2MASS name in format HHMMSSSS$\pm$DDMMSSS\\
      Jmag & float, mag & 2MASS $J$-band magnitude from point source catalogue\\
      e\_Jmag & float, mag & Corresponding error \\
      Hmag & float, mag & 2MASS $H$-band magnitude from point source catalogue\\
      e\_Hmag & float, mag & Corresponding error \\
      Kmag & float, mag & 2MASS $K$-band magnitude from point source catalogue\\
      e\_Kmag & float, mag & Corresponding error \\
      Jext & float, mag & 2MASS $J$-band total magnitude from extended source catalogue\\
      e\_Jext & float, mag & Corresponding error \\
      Hmag & float, mag & 2MASS $H$-band total magnitude from extended source catalogue\\
      e\_Hext & float, mag & Corresponding error \\
      Kext & float, mag & 2MASS $K$-band total magnitude from extended source catalogue\\
      e\_Kext & float, mag & Corresponding error \\
      bj & float, mag & \textit{b$_j$} magnitude from SuperCOSMOS \\
      rf & float, mag & \textit{r$_f$} magnitude from SuperCOSMOS \\
\end{tabular}
\end{table*}

\subsection{Extra Column Description}\label{sec:data:columndescription}
\textbf{type.} A string that indicates the classification of the spectrum. This refers to the Seyfert types at $z<0.5$ (AGN1.0, AGN1.2, AGN1.5, AGN1.8, and AGN1.9), AGN between $0.5<z<1.5$ with Mg{\sc ii} measured (QSOMg), AGN at $z>1.5$ (QSOHiz), and `bad' cases. 

\textbf{ext\_type.} A string that indicates the source exists in an external catalogue. MILLIQUAS entry with a clean spectrum is labelled as MQAGN. Other tags here are Chen18 from \cite{chen18}, 2MAGN from \cite{zaw19}, and 6dAGN from \cite{chen22}.

\textbf{SNR.} Signal-to-Noise ratio of the spectrum close to H$\beta$ or Mg{\sc ii}. Due to the unstable continuum profile and problematic artefacts of 6dFGS spectra, we measure this value with a 20 pixel or $\sim33$\AA\  window at two positions and take the mean. For spectra with $z<0.5$, the two positions around H$\beta$ are measured at 4700\AA\ and 5100\AA. For \textbf{type}=QSOMg spectra, the two positions are 2500\AA\ and 3000\AA. If one value is out of the spectral range or is problematic, then the other value is used instead of taking the mean. SNR for QSO with $z>1.5$ are not measured.
%, since these spectra are {\refbf beyond the intended targets of 6dFGS} and will not be a good representation of the source regardless.

\textbf{FWHM\_X, skew\_X, peak\_X, EW\_X}. Line fitting measurements for H$\beta$ BEL (X=HbB), H$\beta$ NEL (X=HbN), [O{\sc iii}]$\lambda5007$ (X=O3), H$\alpha$ BEL (X=HaB), H$\alpha$ NEL (X=HaN), and Mg{\sc ii} (X=Mg). Narrow emission lines are fitted without a skew and do not have a column for skew. 

{\refbf
The continuum level near the emission lines is required for the calculation of the equivalent width (EW). For H$\beta$ and [O{\sc iii}], this was measured by taking the mean around a 20-pixel window centred around 4700\AA\ and 5100\AA\ respectively (rest wavelength). If 5100\AA\ is out of range for the spectrum, we assume both lines having the same continuum. The same was done for Mg{\sc ii} but for a window around 2500\AA, or 3000\AA\ depending on which is within the spectrum.
}

The full expanded column reads: \textbf{FWHM\_HbB, e\_FWHM\_HbB, skew\_HbB, e\_skew\_HbB, peak\_HbB, e\_peak\_HbB, EW\_HbB, e\_EW\_HbB, FWHM\_HbN, e\_FWHM\_HbN, peak\_HbN, e\_peak\_HbN, EW\_HbN, e\_EW\_HbN, FWHM\_O3, e\_FWHM\_O3, peak\_O3, e\_peak\_O3, EW\_O3, e\_EW\_O3, FWHM\_HaB, e\_FWHM\_HaB, skew\_HaB, e\_skew\_HaB, peak\_HaB, e\_peak\_HaB, EW\_HaB, e\_EW\_HaB, FWHM\_HaN, e\_FWHM, HaN, peak\_HaN, e\_peak\_HaN, EW\_HaN, e\_EW\_HaN, FWHM\_Mg, e\_FWHM\_Mg, skew\_Mg, e\_skew\_Mg, peak\_Mg, e\_peak\_Mg, EW\_Mg, e\_EW\_Mg, 
}

\subsection{Line measurement values and uncertainties}
The emission lines included in the catalogue are the Balmer-NEL, Balmer-BEL, [O{\sc iii}]$\lambda5007$, and Mg{\sc ii}. Both Balmer-BEL lines are mandatory for a catalogue of Seyfert BEL AGN, while [O{\sc iii}]$\lambda$5007 provides the \textbf{r\_HbB\_O3} value for type classification. The Balmer-NEL lines are included for calculations requiring total H$\beta$ and H$\alpha$. 

Each of these lines are fitted to have a sensible uncertainty and to account for any artefacts. However, because Balmer-NEL lines are fitted with a strict upper limit {\refbf in FWHM}, when these lines exceed that limit, the bootstrapping process is unable to accurately estimate the errors. Relaxing the upper limit changes the fitting model and results in a drastically different fit, where the broad component is much smaller. In many cases, this also leads to non-physical or bad fits. As such, the upper limit is retained and the error values for these two lines when they are measured at the FWHM limit are set to -999. 

On the other hand, there are a couple dozen of cases where [O{\sc iii}] exceeds the FWHM$<1200$km s$^{-1}$ limit. For these, the upper limit is relaxed to enable the uncertainty to be calculated. In cases where it is visually obvious that the profile is made up of two components, an additional set of [O{\sc iii}] broad components are included. There are also a handful of cases where [O{\sc iii}] appears much weaker than the noise, preventing a good fit. These are labelled as `bad OIII' in the comments and their uncertainties are set to -999.

In addition, the skews of two-component Gaussian fits are irrelevant and are therefore set to -999. The peak wavelength of two-component fits are set to be the mean of the two peaks, weighted by their respective areas.

\subsection{SMSS photometry and cross-matching}\label{SMSS:matching}
{\refbf
Spectra from 6dFGS are provided in uncalibrated photon counts. In order to estimate the luminosities of measured emission lines, we calibrate the spectra using photometry from the SkyMapper Southern Survey DR4 \citep[SMSS,][]{wolf18,Onken24}. 

All sources in the BEL AGN catalogue lie within 3'' of an SMSS source. In the case of blended sources, most commonly between the AGN host and a star, it is crucial that the photometry is representative of the spectra. Blended sources are identified with the proper motion and object count data from Gaia Data Release 3 \citep{prusti16, vallenari23}. If the SMSS magnitude reflects the measurement of a star from Gaia rather than the AGN, the spectrum of that source will not be calibrated in the catalogue.

In addition, sources with nearby stars will have their 6dF fibre spectra contaminated by starlight, and the provided luminosity estimates will be biased. As a warning, we define a contamination ratio

\begin{align}
    \mathrm{AGN\_to\_STAR} = 10^{-0.4\times M_{AGN} - M_{\mathrm{*}}} ~,
\end{align}

where $M_{AGN}$ and $M_{\mathrm{*}}$ are B$_p$ magnitudes taken from Gaia. The ratio is doubled if the star is further than 3.4'' because it is outside of the 6dFGS 6.7'' aperture and the contamination is assumed to be halved for simplicity. 

Affected targets have the severity noted in the \textbf{Comment} column of the catalogue. We emphasise that this quantity cannot be used to correct the luminosity estimates.
}

\subsection{Luminosity Estimates}\label{sec:data:sub:lumest}
{\refbf 
The catalogue contains line luminosities of the broad H$\beta$ component, the [O{\sc iii}]$\lambda$5007\AA \ line (total flux), and Mg{\sc ii}. The mismatch in scaling between the 6dFGS V and R arm often results in the R arm being scaled down. Since H$\alpha$ always lies within the R arm, the luminosity of the broad H$\alpha$ line cannot be accurately evaluated and is therefore excluded.

Since the spectra are not flux-calibrated, synthetic photometry is not possible. Instead, we calibrate the emission line flux using:

\begin{align}
    F_{\lambda}(x) = \dfrac{\mathrm{counts}(x)}{\mathrm{Val}(ctm(x))} f_{p\lambda}(ctm(x)) = \mathrm{EW}(x)f_{p\lambda}(ctm(x)) \label{eq:lumest}
\end{align}
where $x$ is one of the three emission lines, and $ctm$ stands for the continuum near the emission line (see Section~\ref{sec:data:columndescription} for details). The EW values are provided in the catalogue. $f_{p\lambda}(ctm(x))$ represents the constructed flux values that estimate the calibrated photometric flux for the continuum near an emission line. $F_{\lambda}(x)$ is the calibrated integrated flux, which is subsequently de-reddened using the E(B-V) values from the GNILC dust maps, provided by Planck 2018 \citep{akrami20}, with the Fitzpatrick law \citep{fitzpatrick99} and $a_v=3.1$. The de-reddened fluxes are then converted to luminosities.

The SMSS bands relevant for calibration are $g$ (410-659nm) and $r$ (496-724nm). To construct the required 6.7'' aperture photometry, calibrated magnitudes from images\footnote{The images used were required to have a quality flag {\sc flags < 4} and {\sc nimaflags < 5}.} with 6'' and 8'' aperture were averaged, then linearly interpolated to obtain the 6.7'' aperture magnitude. These values are recorded in the \textbf{mag67\_g} and \textbf{mag67\_r} columns of the catalogue. The pivot wavelengths are $\lambda_{eff, g} = 5100$\AA\ and $\lambda_{eff, r} = 6170$\AA\ \citep{bessell11, tonry18}. The corresponding fluxes are given in the \textbf{flux67\_g} and \textbf{flux67\_r} columns. Associated uncertainties for these four columns are included, calculated using the standard deviation. For all sources in the catalogue, the percentage error in these values are $<3\%$, with an average of $0.3\%$.

Next, consider the ideal case of a featureless linear continuum. Now, \textbf{flux67\_g} and \textbf{flux67\_r} represent the flux at their respective pivot wavelengths. The flux at any wavelength, $f_{p\lambda}(ctm(x))$, can then be determined by interpolating or extrapolating between these two flux values.

In general, spectra do not have linear continua, which causes the accuracy of the interpolation or extrapolation to decrease with distance from the pivot wavelength. One method to avoid this would be to scale the photon counts at the pivot wavelengths. Theoretically, this would be a more accurate method because it follows the spectral shape. However, the spectral shape of 6dFGS is uncorrected, particularly at longer wavelengths. Therefore, the interpolation/extrapolation method is preferred as it acts as a correction.

Once $f_{p\lambda}(ctm(x))$ is constructed and applied in Eq~\ref{eq:lumest}, estimating the luminosity is straightforward. The errors in the luminosity arise from the uncertainty in the SMSS magnitude and our measured EW. The EW is the main source of error and its scale depends on the 6dFGS data quality. For the H$\beta$ BEL, the median of this error is $\sim$10\%, while the maximum is $\sim150\%$. 

For completeness, the linear interpolation for the 6.7'' aperture magnitude assumes that the galaxy surface brightness curve is linear. Together with the assumption of a linear continuum, these steps introduce systematic errors that are generally difficult to quantify. A more accurate approach for modelling the brightness curve would be to use a low-order polynomial or spline over the 5'', 6'', 8'' and 10'' apertures. However, testing this on 100 sources revealed no significant differences. Meanwhile, obtaining $f_{p\lambda}(ctm(x))$ by scaling rather than interpolating results in an average difference of 6\% in the estimated luminosity.
}

\section{Discussion}
\subsection{Region of Completeness}\label{sec:discussion:sub:completeness}
6dFGS, and subsequently this BEL AGN catalogue, can be confounding to use in practical cases because of the two sets of survey targets that originate from a wide array of samples with different properties. In this section, we explain the use of our catalogue to select  a sky-complete BEL AGN sample. 

\begin{figure*}
    \centering
    \includegraphics[width=0.42\textwidth]{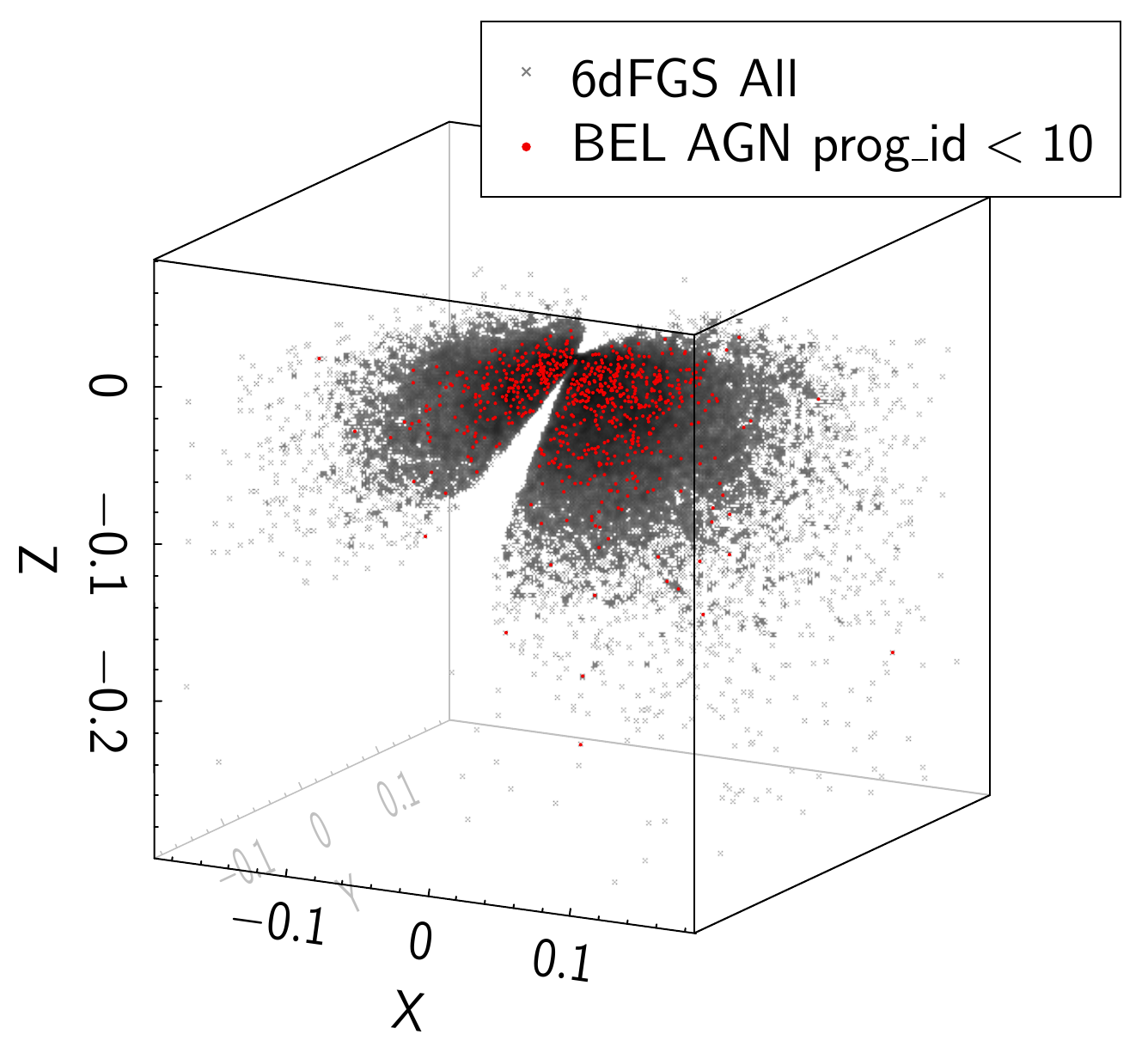}
    \includegraphics[width=0.42\textwidth]{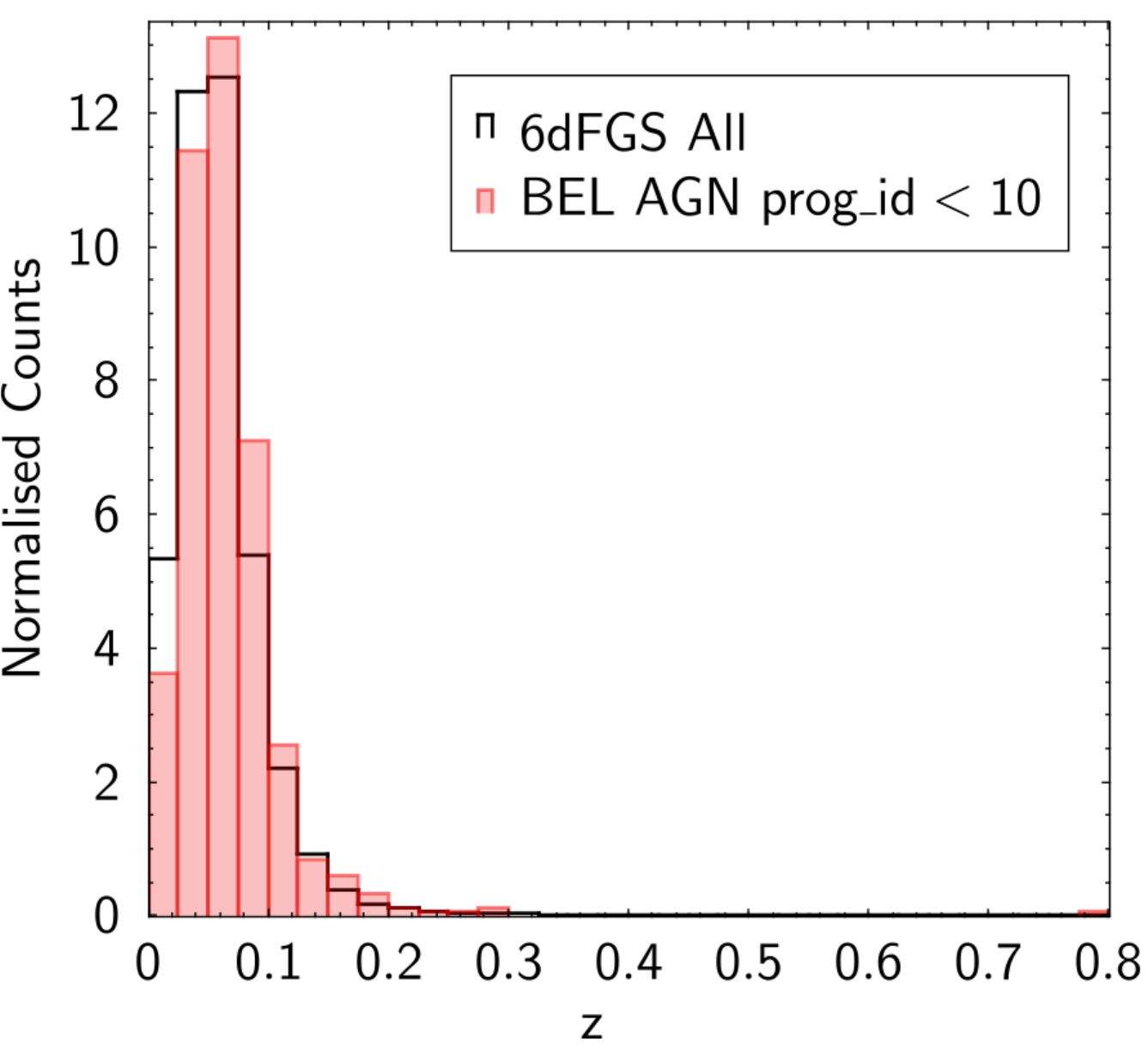}
    \includegraphics[width=0.97\textwidth]{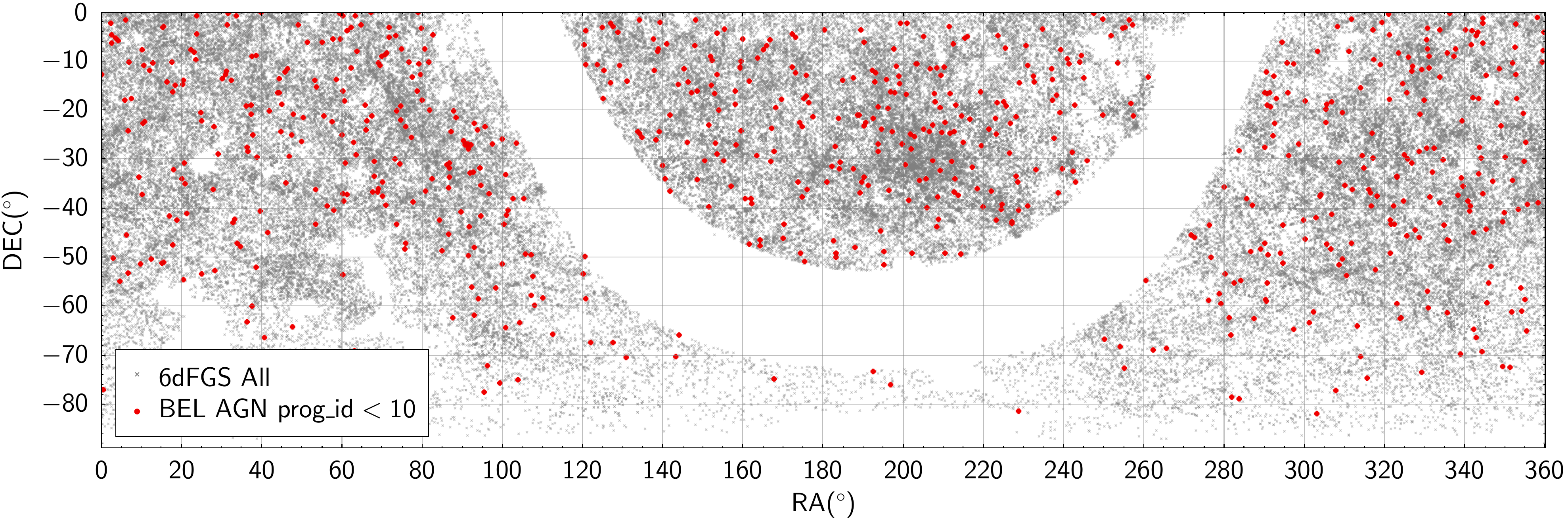}
    \caption{Figures demonstrating the similarity of distribution in volume, redshift, and sky coverage between all of 6dFGS and the BEL AGN with \textbf{prog\_id} $<10$. {\sl Top left:} volume coverage; {\sl Top right:} redshift distribution; {\sl Bottom:} is the sky coverage.}
    \label{fig:chp:6dfagn:main_cov}
\end{figure*}

{\refbf Assuming that our sample of BEL AGN is catalogue-complete, the sky-completeness therefore follows the same distribution and flux statistics as the 6dFGS sources.} Since only the extended sources in the 6dFGS flux limited sample is sky-complete in flux, one can only form a list of BEL AGN that is also sky-complete in flux. This is done by selecting for sources with \textbf{prog\_id} between 1 to 8 (refer to as $<10$) and AGN from Type-1.0 to 1.9 in \textbf{type}, for a sample of 665 AGN (including 12 AGN that are highly contaminated by neighbouring star). Figure~\ref{fig:chp:6dfagn:main_cov} shows that this subset of our catalogue does indeed follow the density and distribution of sources in 6dFGS as expected.

{\refbf To verify the assumption of catalogue-completeness, our catalogue is compared to two other smaller 6dFGS AGN samples. The catalogue of narrow line Seyfert-1 within 6dFGS created by \cite[][hereafter Chen18]{chen18} contains 167 sources. All of these sources are recovered in our catalogue. 

The second catalogue is the recent} Swift/BAT AGN spectroscopy survey \citep[BASS,][]{koss17}. Their data release 2 \citep{koss22} contains 858 sources from \cite{koss2022vizier} and there are 187 common sources with 6dFGS within a 10'' radius. 102 of them are also in the BEL AGN catalogue. Theremaining  85 BASS sources are made up of 71 Type-2, 3 blazars with no emission lines, 9 Type-1.9, and 2 Type-1.0. The Type-2s and blazars are irrelevant to our catalogue, and below we argue that the Type-1.9 and Type-1.0's do not meet our selection criteria. 

We inspect these BASS spectrum and line fitting results from \url{https://www.bass-survey.com/}. One of the BASS Type-1.0 is 6dFGS g0519358-323928. This is a Changing-Look AGN where the BEL only emerged in recent years \citep{hon2022}, which means that 6dFGS observed it as a Type-2 and it does not belong in the BEL AGN catalogue. Meanwhile, the other BASS Type-1.0 and six of the Type-1.9s were rejected because they do not satisfy our criteria for a BEL listed in Section~\ref{sec:methods:sub:lowzbel}. Then there are two BASS Type-1.9 that appear to be a mistyping from BASS as their line fitting results show no broad components. The last one is clearly a Type-1.9 in the BASS spectrum, but is not possible to see this in the 6dFGS spectrum due to the instrument artefacts around H$\alpha$. As such these spectra do not make it into our catalogue. 

{\refbf
Therefore, when comparing to Chen18 and BASS, the BEL AGN sample is 100\% catalogue-complete in the context of selection based on 6dFGS spectra. In a broader context, where the selection is based on all available spectra and pre-existing classifications, our selection does not include 8 BASS sources (excluding the two mistyping cases and one Changing-Look). The catalogue-completeness in this context is then 93\%.

BASS sources rejected by our criteria as BEL are the main source of this discrepancy in catalogue-completeness, and Type-1.8s and 1.9s are mostly affected. This is because there is a single BEL  and H$\alpha$ tends to appear weaker compared to the H$\alpha$ of Type-1.2s or 1.0s. Relaxing the criteria for BEL to recover these sources is not desirable as it will introduce non-BEL AGN contaminants into our catalogue. 

Regardless, if we cannot clearly `see' the BEL, it implies that the quality of the spectra is not reliable and a definitive classification is not possible.
} 

\subsection{Reliability of measurements}\label{sec:discussion:sub:reliability}
\begin{figure*}
    \centering
    \includegraphics[width=0.4\textwidth]{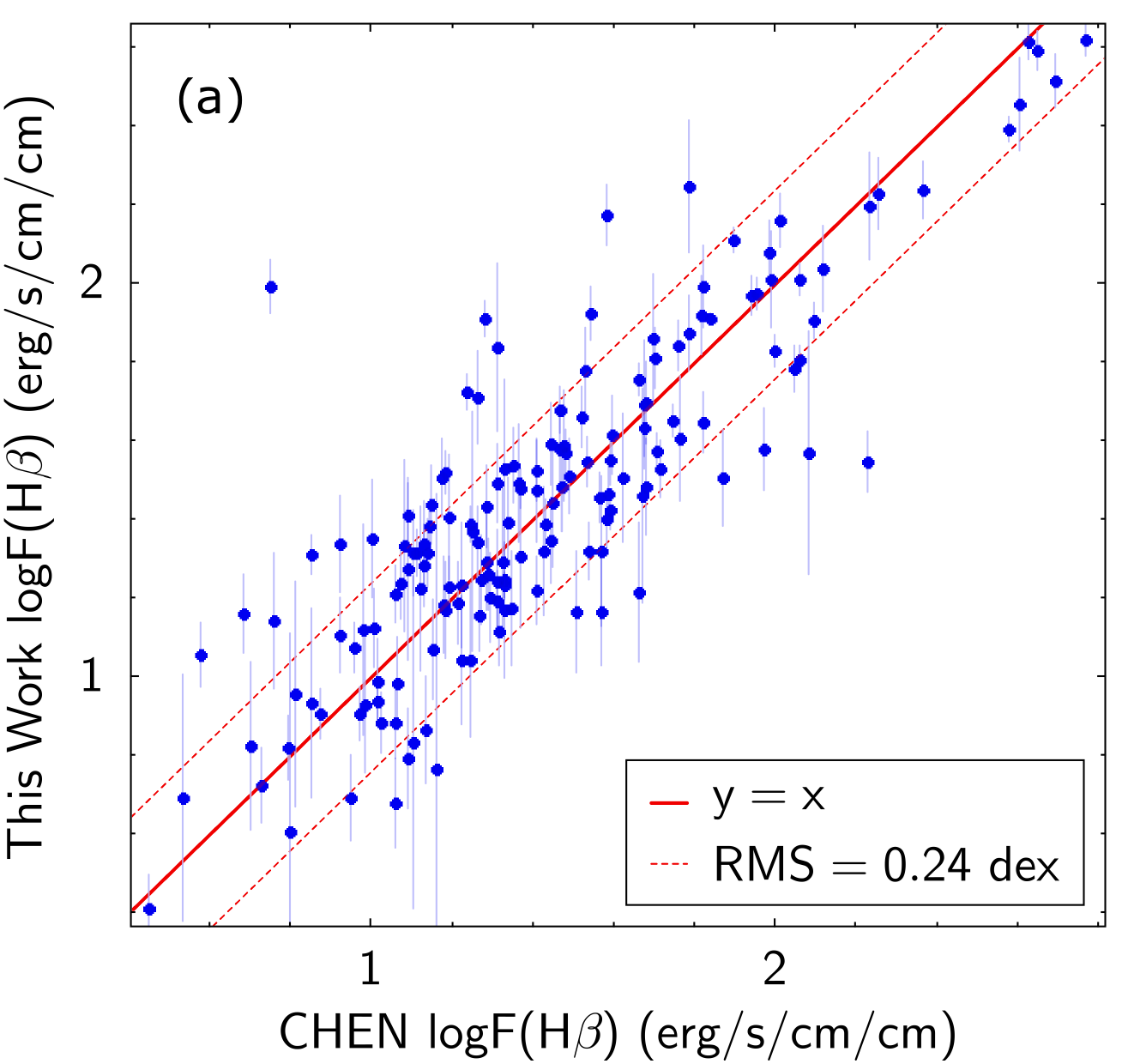}
    \includegraphics[width=0.4\textwidth]{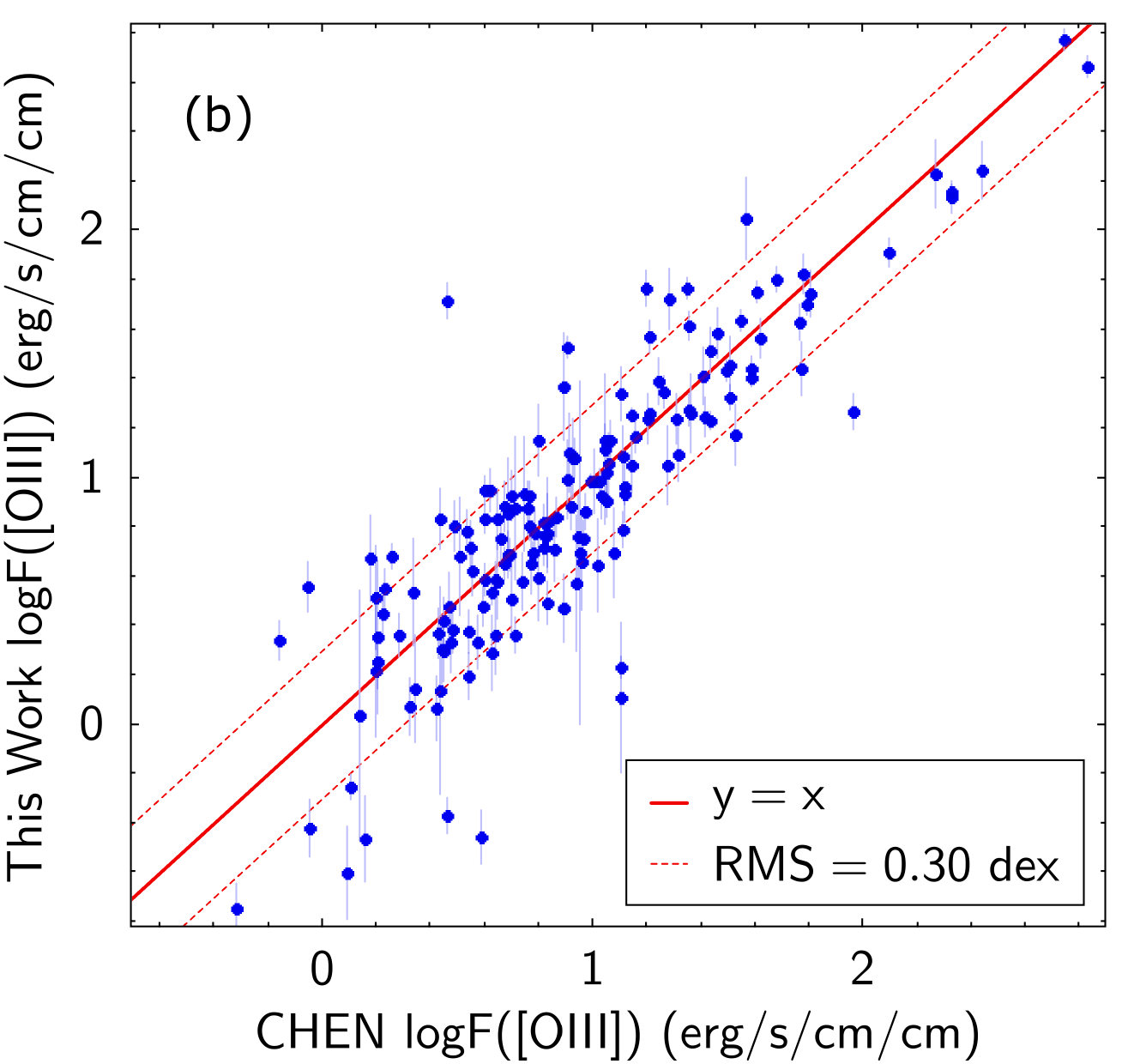}
    \caption{{\refbf Plots comparing our measurements to Chen18 and BASS. In (a) and (b), we compare our total H$\beta$ flux and [O{\sc iii}] with Chen18's. In each, the linear relation $y=x$ is shown in red, and the best fit to the data is shown in grey. The dotted lines are the RMS with respect to the linear relation}}
    \label{fig:chp:6dfagn:chen_compare}
\end{figure*}

\begin{figure}
    \centering
    \includegraphics[width=0.92\linewidth]{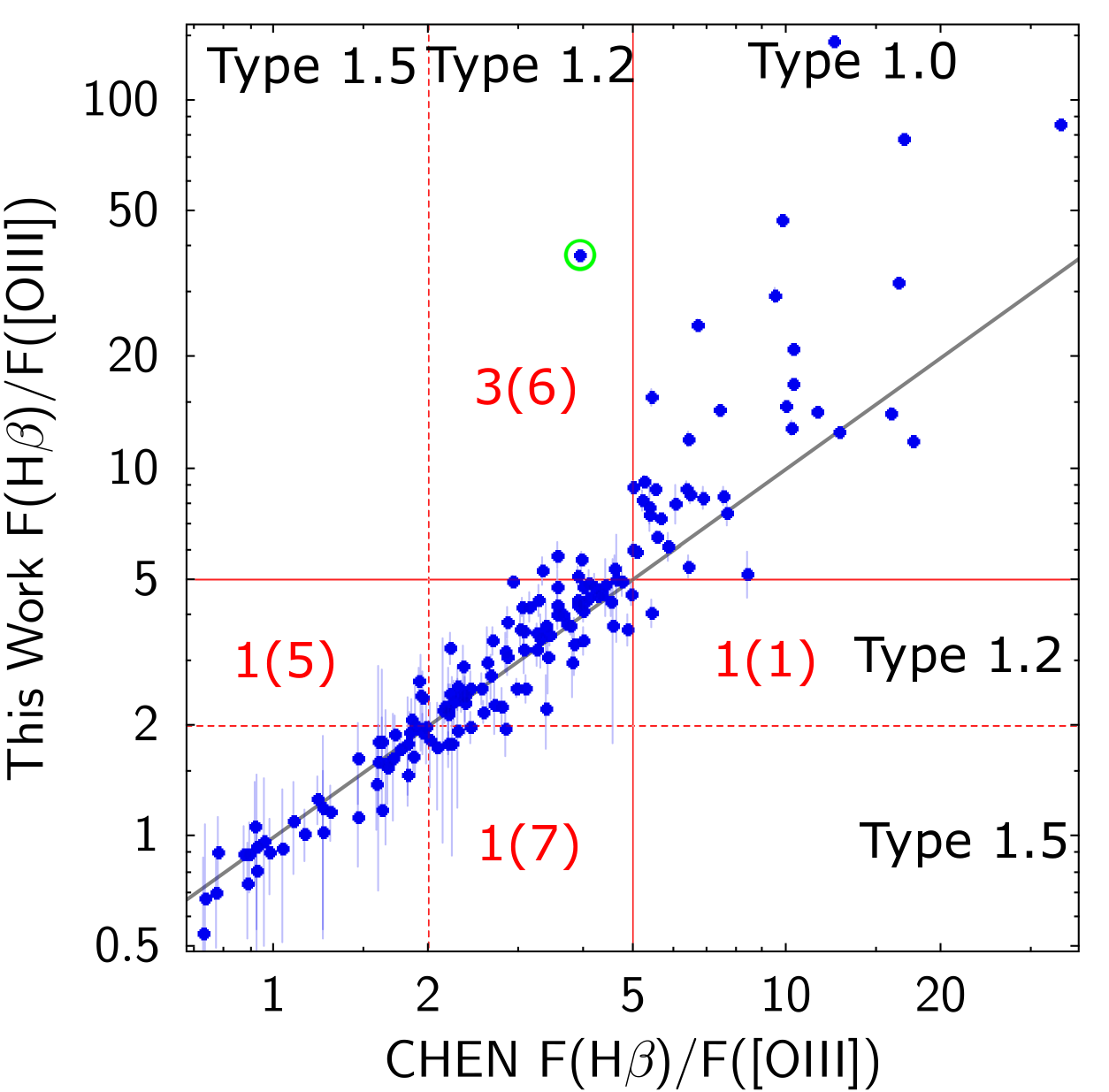}
    \caption{{\refbf Plot comparing the total H$\beta$ (BEL + NEL) and [O{\sc iii}] integrated flux ratio between our work and from CHEN. Grey solid line indicates the $y=x$ linear relation. Dotted vertical and horizontal lines indicates the boundary between Type-1.5 and 1.2. Solid vertical and horizontal lines indicates the boundary between Type-1.2 and 1.0. Red text is the count of objects with disagreeing type accounting for uncertainty, and the value in bracket is the count without accounting for uncertainty. The object marked with a green circle is the outlier mentioned in the text}}
    \label{fig:chp:6dfagn:chen_R}
\end{figure}

{\refbf
This section assesses the reliability of our flux calibration by comparing them to Chen18. In Chen18, measured 6dFGS fluxes were calibrated by utilising seven sources with SDSS spectra. Using both SDSS and 6dFGS spectra, Chen18 obtained a sensitivity curve over wavelength for a general shape correction, and a curve of average flux/count ratio for normalisation. This calibration was then verified by comparing the relation between the continuum flux and the optical magnitude of the re-calibrated 6dFGS spectra to another SDSS sample of narrow line Seyfert-1.

However, because Chen18 is specifically interested in narrow Line Seyfert-1s, they fit their lines with three Gaussian components and the total profile includes the narrow H$\beta$ line. In order to minimise this disparity, when comparing H$\beta$, the broad and narrow components from our measurements are added together. 

The comparison is shown in Figure~\ref{fig:chp:6dfagn:chen_compare}a-b for H$\beta$ and [O{\sc iii}]. For both lines, the agreement with Chen18 is statistically high with a Pearson correlation coefficient squared (r$^2$) of 0.72 and 0.75 respectively. The measurements have a scatter (relative to linear relationship) of 0.24 and 0.30 dex. This corresponds to a difference with a factor that is $<2$ for 68\% of objects. These statistics are of two sets of values from different measurements and calibration method. With that in consideration, the similarity to Chen18 suggests that our method is sufficiently reliable for the purpose of estimating line luminosity.

Since there are discrepancies in flux, there will also be discrepancies in the R ratio and therefore the AGN type classification. Figure~\ref{fig:chp:6dfagn:chen_R} plots the ratio as measured by Chen18 and from our work, showing minor variations in value among Type-1.5 and 1.2, but large discrepancies for Type-1.0 objects. Among the 160 sources plotted, 19 ($\sim12\%$) have a disagreement in type. A simple way to account for the uncertainty would be to assume agreement if the error bars cross the type boundary. This leaves only 6 ($\sim4\%$) sources with a disagreement in type.

Regarding the Type-1.0 objects with large discrepancies in R ratio, these are sources where the amplitude of the Fe{\sc ii} emission exceeds the peak flux of [O{\sc iii}]$\lambda5007$. The outlier source marked with a green circle in Figure~\ref{fig:chp:6dfagn:chen_R} is an example of such case. This source is g0400244-250444 and our R ratio is ten times greater than Chen18. The weak contrast between [O{\sc iii}]$\lambda5007$ and Fe{\sc ii} emission exaggerates the different Fe multiplet subtraction methods during continuum fitting, causing our measurements to have a low integrated flux for [O{\sc iii}]$\lambda5007$, and a large R ratio. 

This comparison suggests a strong similarity to Chen18, with minor disagreements in AGN classification for only $\sim10\%$ of objects. In general, most discrepancies in the R ratio is insufficient to cause a change in AGN type. Regardless of these disagreements, these sources have been identified as BEL AGN in our catalogue, which is the main purpose of this work.

}

\subsection{Characteristics of the full catalogue}\label{sec:discussion:sub:charac}

{\refbf Users of this catalogue need to be aware that the parent sample of 6dFGS spectra is a flux-limited, nearly hemispheric galaxy sample with the addition of auxiliary targets. These additional targets might be fainter or be drawn from other multi-wavelength catalogues. The flux-limited sample (defined by \textbf{prog\_id} $< 10$) contains low-luminosity AGN out to a redshift of $\sim 0.2$, while the auxiliary targets (\textbf{prog\_id} $>10$) are a highly incomplete list of quasars up to $z\approx 3.8$. This section illustrates the characteristics of targets from both categories in our catalogue.
}

\begin{figure*}
    \centering
    \includegraphics[width=0.99\textwidth]{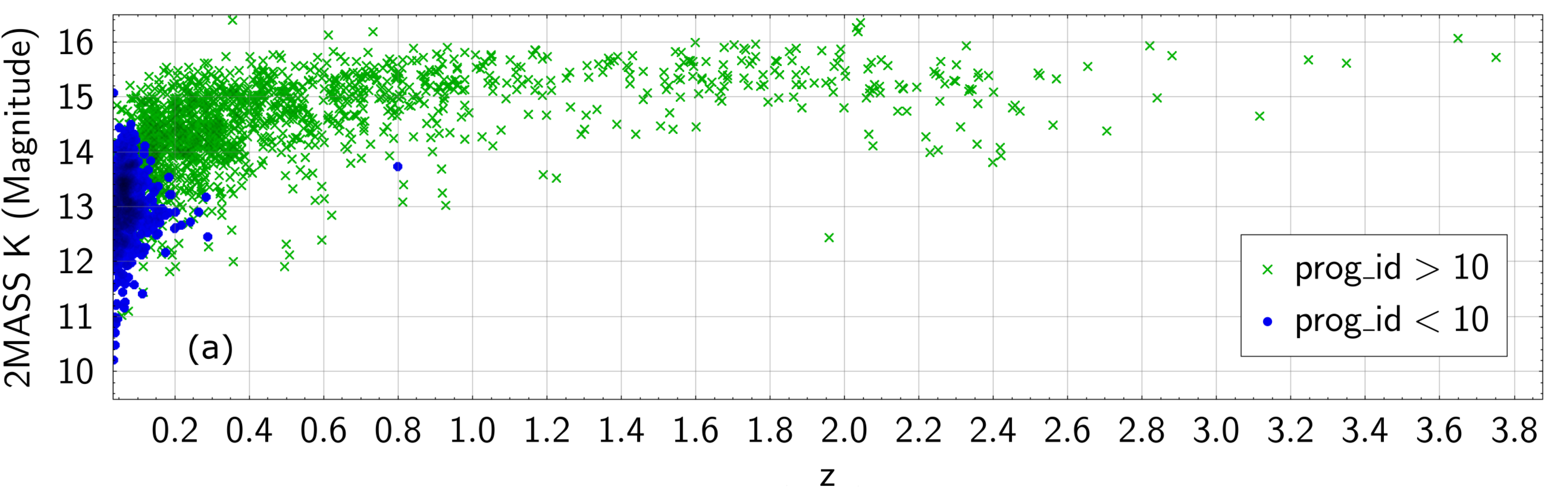}
    \includegraphics[width=0.49\textwidth]{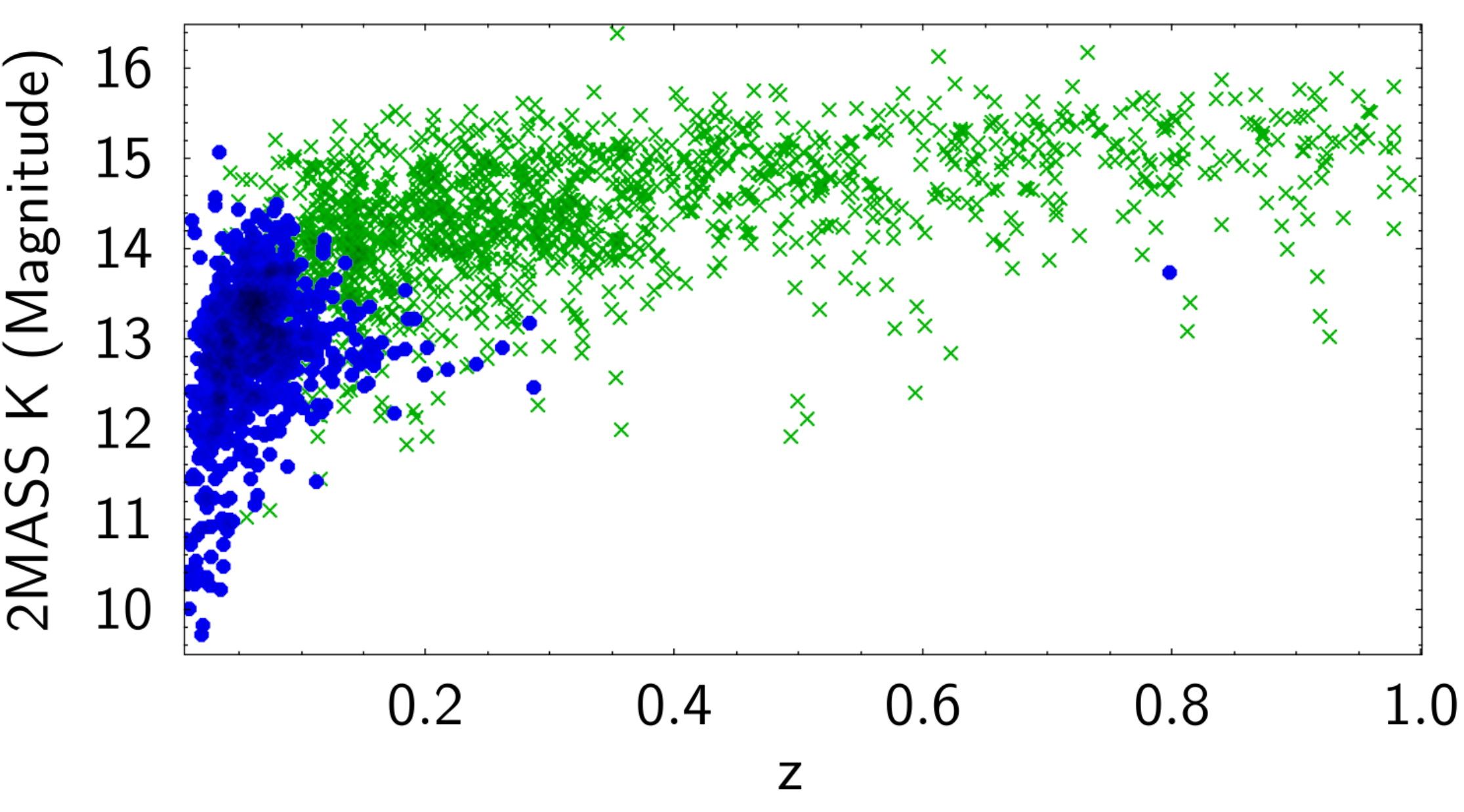}
    \includegraphics[width=0.49\textwidth]{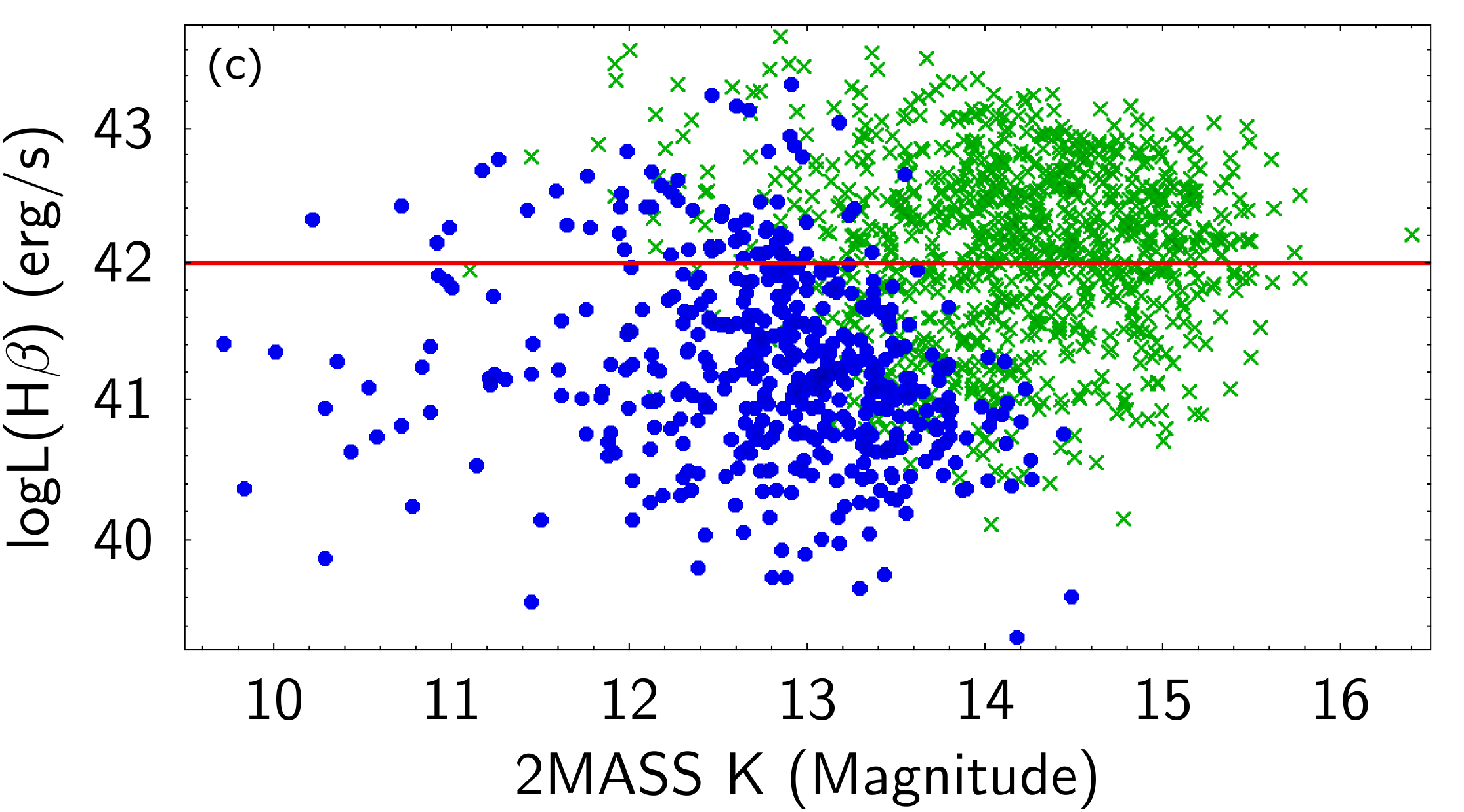}
    \includegraphics[width=0.49\textwidth]{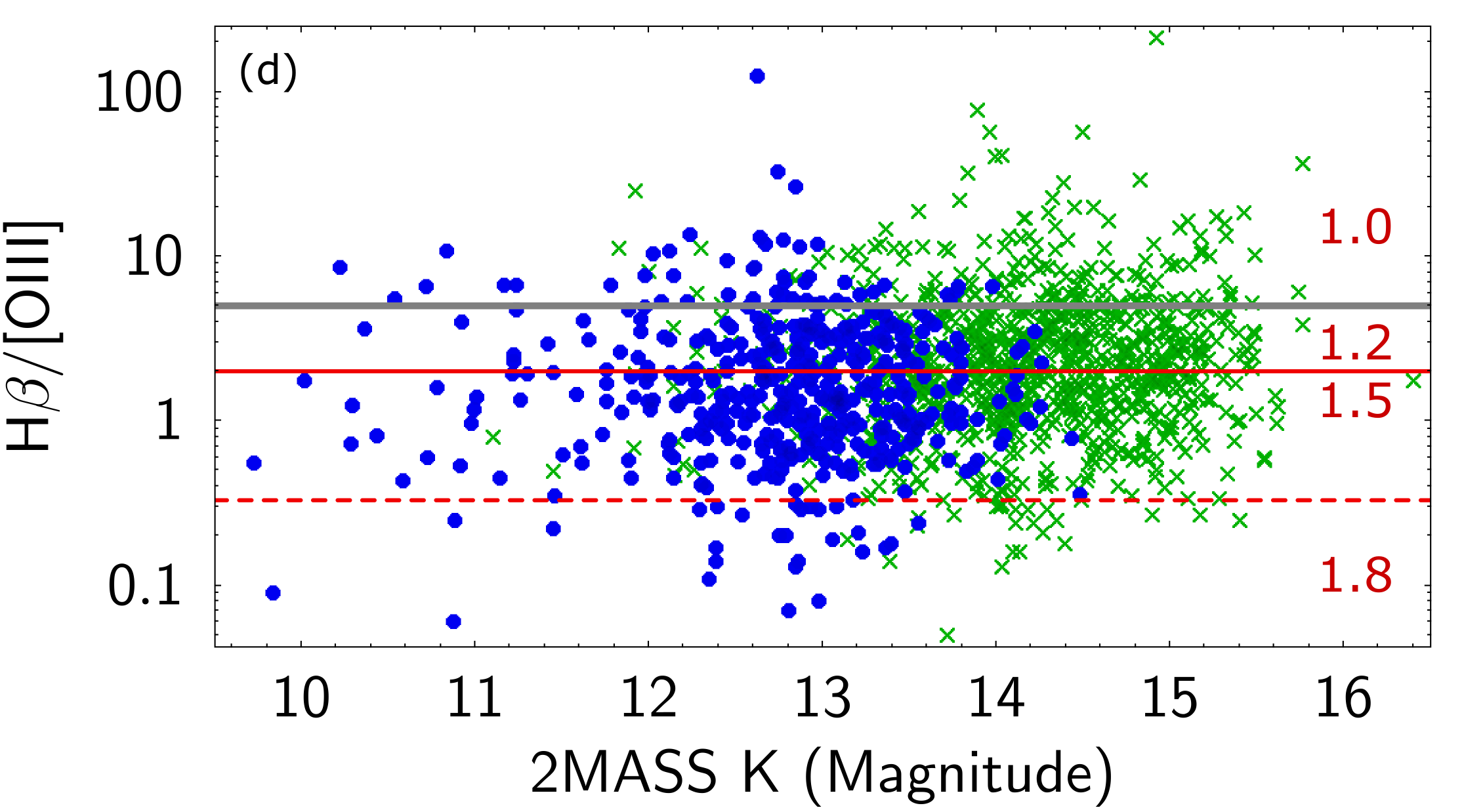}
    \includegraphics[width=0.49\textwidth]{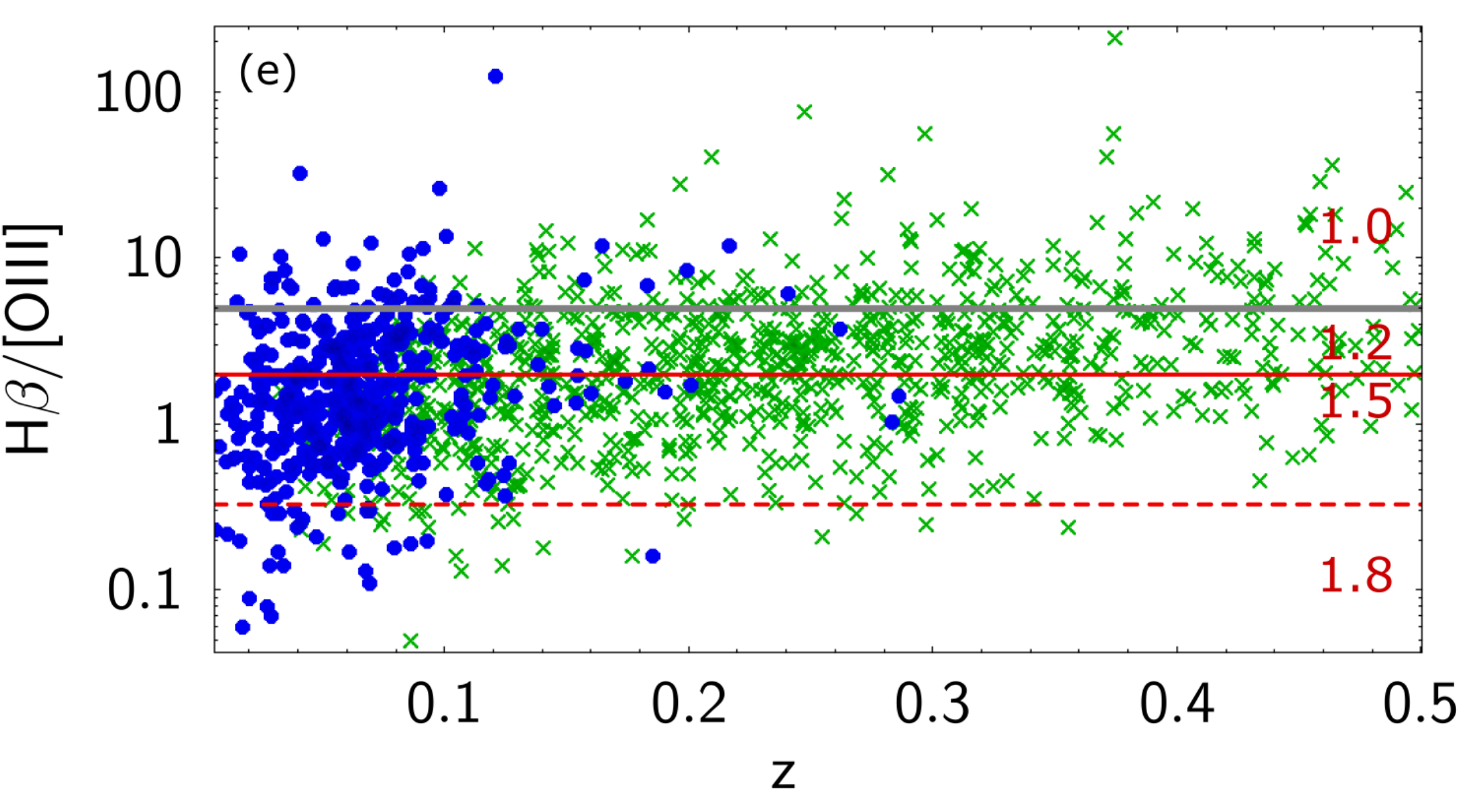}
    \includegraphics[width=0.49\textwidth]{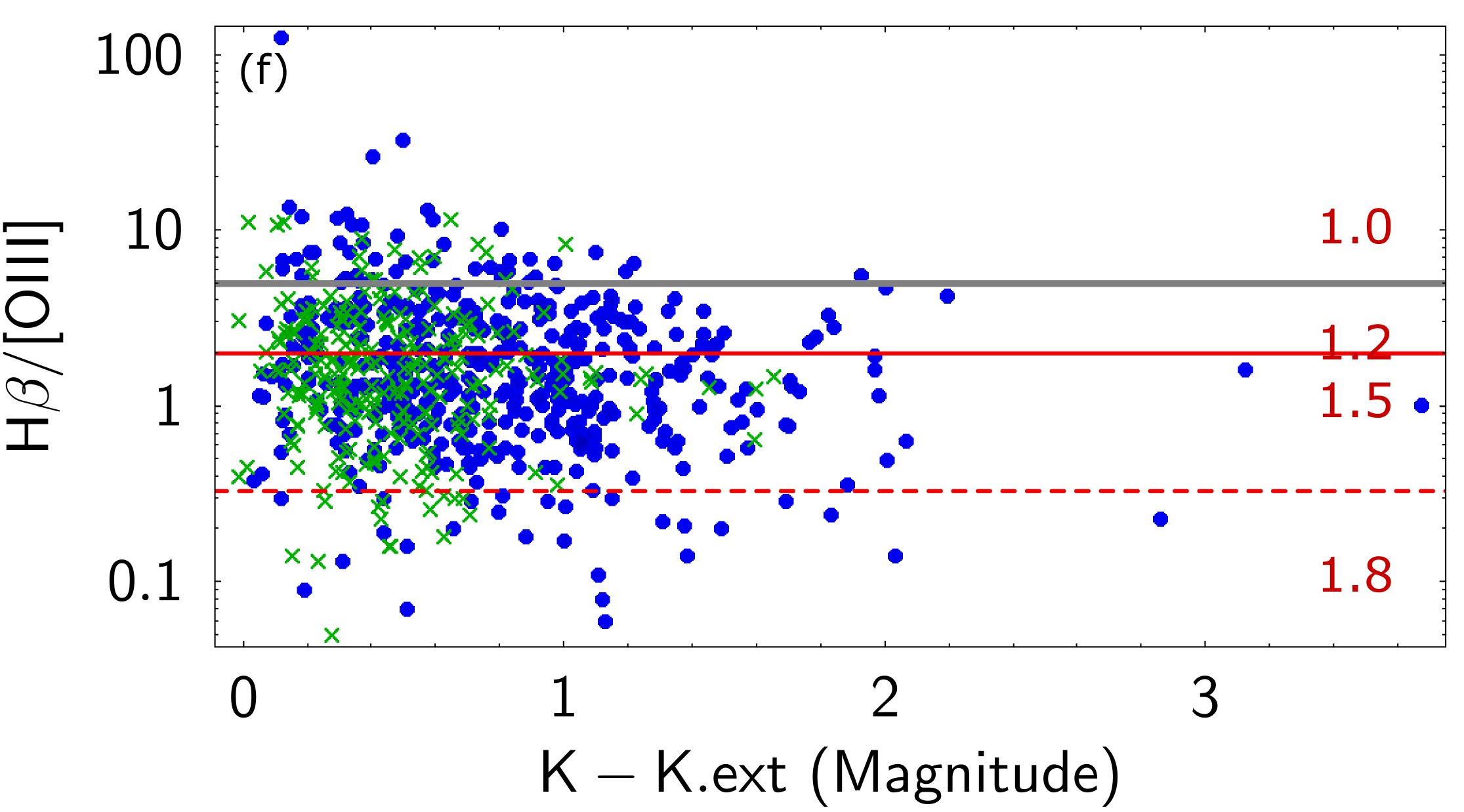}
    \includegraphics[width=0.49\textwidth]{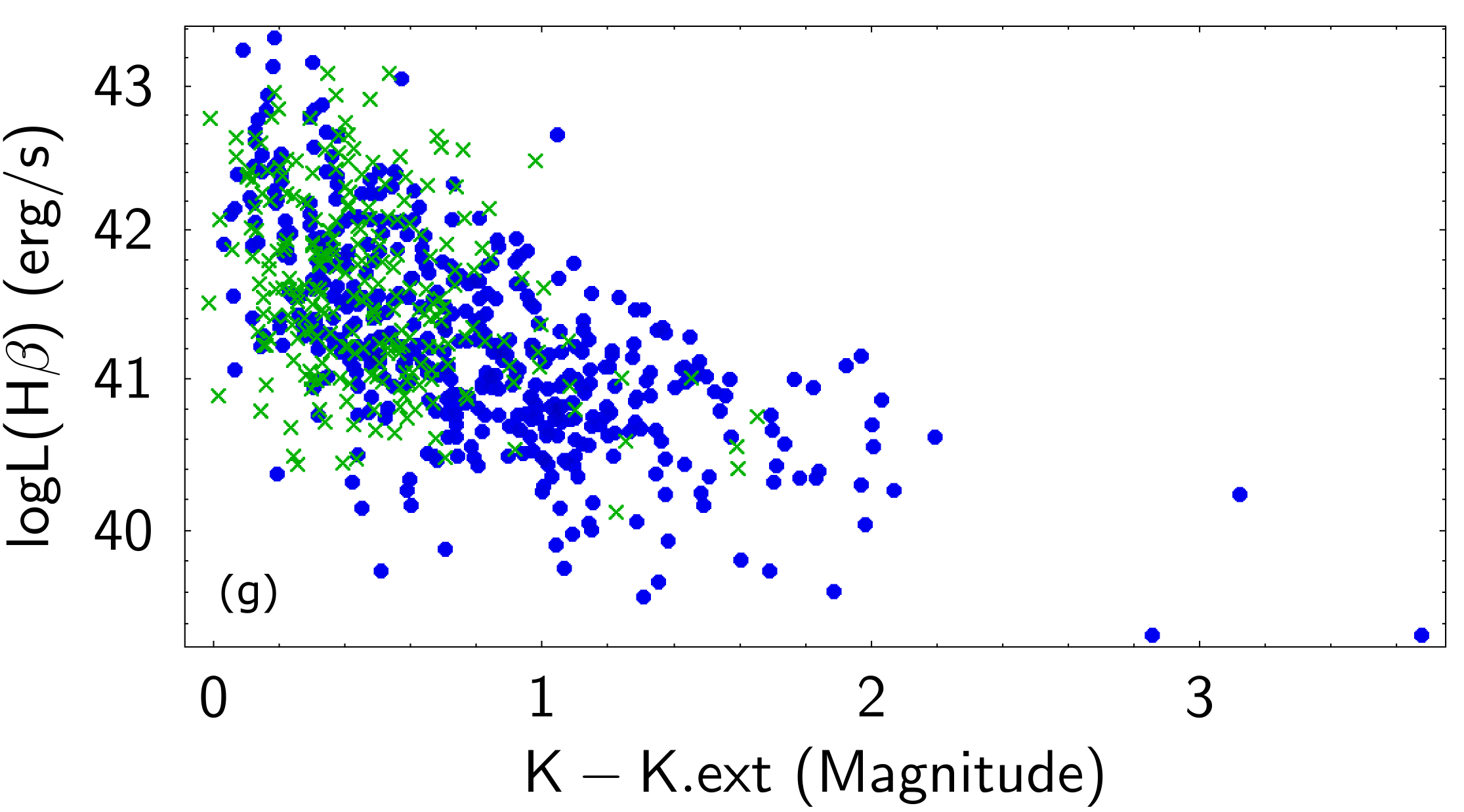}
    \caption{Figures comparing the distribution of flux limited and the auxiliary targets. In all of these, the flux limited targets (\textbf{prog\_id} $< 10$) are represented with filled circles and the auxiliary targets (\textbf{prog\_id} $> 10$) with a plus. Plot (a) shows the catalogue at the full redshift range, while plot (b) goes up to $z=1$ to show the flux limited targets clearly. Plot (c) shows the distribution of line luminosity in relation to apparent magnitude. Plot (d) is the same as plot (c) but with R-value as the y-axis. Plot (e) shows the distribution of R-value in relation to redshift. The horizontal lines in plot (d), (e), (f) indicates the boundary for Type-1.8, Type-1.5, and Type-1.2. Plot (f), (g) shows the distribution of R-value and line luminosity in relation to source apparent size (approximated with 2MASS photometry Kmag-Kext, see text and Table~\ref{tab:chp:6dfagn:colname}).}
    \label{fig:chp:6dfagn:character}
\end{figure*}

The flux-limited sample occupies a low-redshift region below $z\sim 0.2$ with a mean of 12.9 mag in $K_{2MASS}$ (see Figure~\ref{fig:chp:6dfagn:character}a and Figure~\ref{fig:chp:6dfagn:character}b). The auxiliary targets extends this redshift distribution above $z\sim 0.2$ around a mean of $K_{2MASS}=13.9$ magnitude, but only contains the luminous sources above 16 magnitude at all redshifts.

Due to this luminosity bias and the redshift scaling of apparent magnitude, the majority of the high luminosity AGN in our catalogue are the auxiliary targets (see Figure~\ref{fig:chp:6dfagn:character}c). We also see from Figure~\ref{fig:chp:6dfagn:character}d that AGN are equally distributed in apparent brightness (around a mean of 13 to 14 mag in 2MASS K) irrespective of AGN type. This is expected since the AGN type is correlated with luminosity and not apparent magnitude. 

{\refbf Figure~\ref{fig:chp:6dfagn:character}e shows that there are fewer Type-1.8 and more Type-1 AGN as redshift increases. This is an effect of instrument limitation where the fainter H$\beta$ flux at higher redshift is difficult to detect above the noise.}

Point source magnitude K subtracted with extended source magnitude K is a proxy for apparent size. Looking at their distribution in Figure~\ref{fig:chp:6dfagn:character}f, Figure~\ref{fig:chp:6dfagn:character}g, we note that Type-1 and high luminosity AGN are exclusively point-like objects (K - K.ext $\sim0.5$). Auxiliary targets are also all point-like, which is expected since extended sources are already all covered in the flux limited survey. There appears to be a trend with H$\beta$ luminosity based on apparent size, where the larger the apparent size the dimmer the AGN. This applies to the type classification, where larger apparent sizes have weaker AGN types. These are very likely due to host galaxy contamination of the spectra, causing the H$\beta$ BEL relative strength over the galaxy to be underestimated especially because these spectra are observed with a wide 6.7'' aperture.

\subsection{AGN new to literature}\label{sec:discussion:sub:newtolit}
To quantify how many AGN in our catalogue are new to the literature, we compare to the newest version of MILLIQUAS at the time of writing this paper, Version 8 (22 July 2023). This version contains AGN from notable surveys such as SDSS, BASS, LAMOST, UVQS and a compilation of papers presenting optical spectra. However, it does not contain the sources from Chen18 \citep{chen18}. By combining all matches with MILLIQUAS and Chen18, we find 1719 AGN in our catalogue that has been previously published in the literature. 

\begin{figure*}
    \centering
    \includegraphics[width=0.92\textwidth]{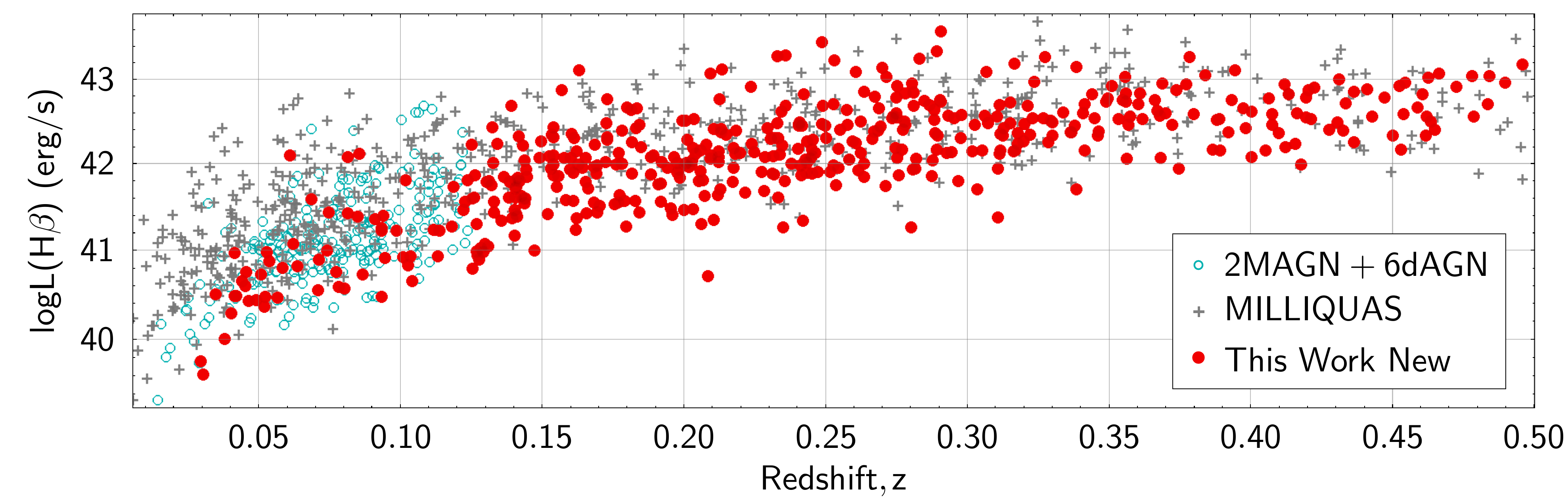}\vspace{-2em}
    \includegraphics[width=0.92\textwidth]{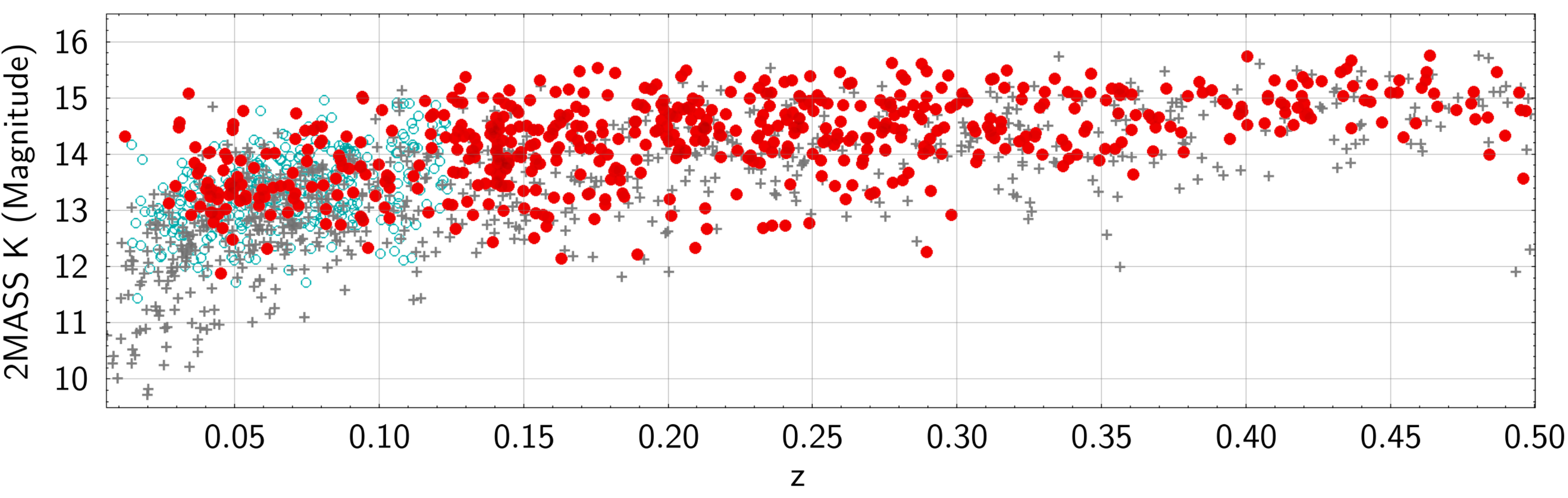}
    \caption{Figures showing the H$\beta$ luminosity and 2MASS K-magnitude distribution of sources in our catalogue, separated by colours to demonstrate the differences between those new to the literature, and those that have been previously known.}
    \label{fig:chp:6dfagn:newtolit}
\end{figure*}

This leaves 578 Type-1.0 to 1.9 AGN at $z<0.5$ and 313 AGN (`Mg' and `QSO') at $z>0.5$. This is 891 AGN that are identified for the first time in the literature  $\sim15$ years after the completion of 6dFGS. On top of this, there are also 79 `bad' spectra that we plan to follow up to verify their classification. In Figure~\ref{fig:chp:6dfagn:newtolit}, we show that most of the `new' AGN classified in this work are lower in luminosity and magnitude compared to those already known in MILLIQUAS. Therefore, this work completes the lower end of the AGN population within 6dFGS.

\subsection{Comparison to 2MAGN and 6dAGN}\label{sec:discussion:sub:2magn}
{\refbf
The MILLIQUAS reference codes 2MAGN and 6dAGN refers to works conducted by the same group of authors, \cite{zaw19} and \cite{chen22}. These authors have attempted to classify 6dFGS spectra for $z<0.13$. Their catalogue has two main categories: Sy1, and narrow line AGN. Sy1 includes 3\,109 BEL AGN, indicated from the presence of a broad H$\alpha$. The latter category consists of 12\,156 sources selected using a BPT diagram. Both of these works employed highly automated algorithms, and we noted many discrepancies when comparing their list of Sy1 to our catalogue. For simplicity, we will group and refer to both works as 2MAGN.

\subsubsection{Common sources}
There are 952 AGN with $z<0.13$ in our catalogue under any label in the \textbf{type} column. However, only 678 of these are also in the list of Sy1 of 2MAGN. The disagreement suggests that the two catalogues represent distinct subsets of the true BEL AGN population in 6dFGS, with only a moderate degree of overlap.

Since the procedure in 2MAGN is automated, it is not sensitive to the issues related to 6dFGS discussed in this work. The catalogue lacks indicators for incorrect 6dFGS redshifts (Figure~\ref{fig:chp:6dfagn:chenex_bad}a), non-physical continua (Figure~\ref{fig:chp:6dfagn:chenex_bad}b), cross-talk and ISM contamination, as well as other issues that prevent broad profile identification, such as fringing effects and telluric absorptions (Figure~\ref{fig:chp:6dfagn:chenex_bad}c-e).

Among the 678 common sources, we have labelled 18 spectra as 'bad' (see end of Section~\ref{fitting_problems}), 41 are flagged with telluric issues, 8 with major star contribution, 10 with ISM or other fibre contaminations, 1 case of cross-talk, 1 case of incorrect 6dFGS redshift, 2 with fringes preventing identification, and 4 additional cases with other issues. Some of these issues, like the non-physical continuum in Figure~\ref{fig:chp:6dfagn:chenex_bad}b, do not prevent identification but are detrimental to modelling the H$\beta$ BEL reliably. Others completely prevent identification of the BEL. Flagging these cases accordingly is essential for the usability of the catalogue. 

\subsubsection{2MAGN sources not in our catalogue}
There are 2\,431 2MAGN sources not included in our BEL AGN list. After visually inspecting all of them, there are only two that can be classified as Type-1.9s and missed by our algorithm in Section~\ref{sec:method:sub:RFfinder}. Both of these are now included in our catalogue to achieve close to 100\% catalogue-completeness.

If we relax the BEL criteria described in Section~\ref{sec:methods:sub:lowzbel}, there may be additional sources from 2MAGN that we have missed. Using the 93\% value for catalogue-completeness found in Section~\ref{sec:discussion:sub:completeness}, the theoretical number of common sources between 2MAGN and this work should increase to $\sim730$, suggesting a loss of only $\sim50$ Type-1.9s compared to the actual value of 678 common sources. The remaining $\sim2000$ unaccounted Sy1 in 2MAGN are likely misclassified. 

The spectra in Figure~\ref{fig:chp:6dfagn:chenex_bad}f-h are examples of these misclassified sources. It is also important to point out that these spectra do not become BEL AGN with a different line fitting method. Some represent fundamental spectral errors (i.e. fibre contamination or fringing) that 2MAGN has not accounted for, regardless of signal-to-noise ratio. 

\subsubsection{BEL AGN sources not in 2MAGN}
There are also 230 BEL AGN that are not classified as Sy1 in 2MAGN. Some of these sources are instead classified as Sy2 or composite galaxies. A majority of them are dimmer sources that are easily missed by an automated algorithm, as identifying them often requires a human eye. However, we find that even spectra of AGN with reasonable quality and strong broad emission lines are overlooked by 2MAGN as illustrated by the example shown in Figure~\ref{fig:chp:6dfagn:chenex_bad}i.

\subsubsection{Measurements}
Figure~\ref{fig:chen-fwhm} compares the FWHM measurements between 2MAGN and our work. Our FWHM measurements are consistently broader than those in 2MAGN, with a few outliers where the FWHM in 2MAGN is significantly broader. 

Two examples are presented in Figure~\ref{fig:chp:6dfagn:chenex-fwhm}a-d. The FWHM measured for g0547318-304146 by us is 3\,900 km/s from H$\beta$, while 2MAGN measures 17\,700 km/s. For g1328311-490906, we measure 5\,500 km/s, whereas 2MAGN finds 28\,000 km/s. In these two cases, there are no issues with our line fitting and the residual is consistent around the zero level. 

Conversely, 2MAGN sometimes reports a much lower FWHM value. For g1254564-265702, which has a very broad H$\alpha$ (Figure~\ref{fig:chp:6dfagn:chenex-fwhm}e), we measure a FWHM of 8\,200 km/s, while 2MAGN reports 3\,500 km/s.  For g1928421-251356, the broad H$\alpha$ exhibits a double-peaked profile (Figure~\ref{fig:chp:6dfagn:chenex-fwhm}f), with our measurement of 9\,000 km/s compared to 2\,200 km/s from 2MAGN.

\begin{figure*}
    \centering
    \includegraphics[width=0.85\textwidth]{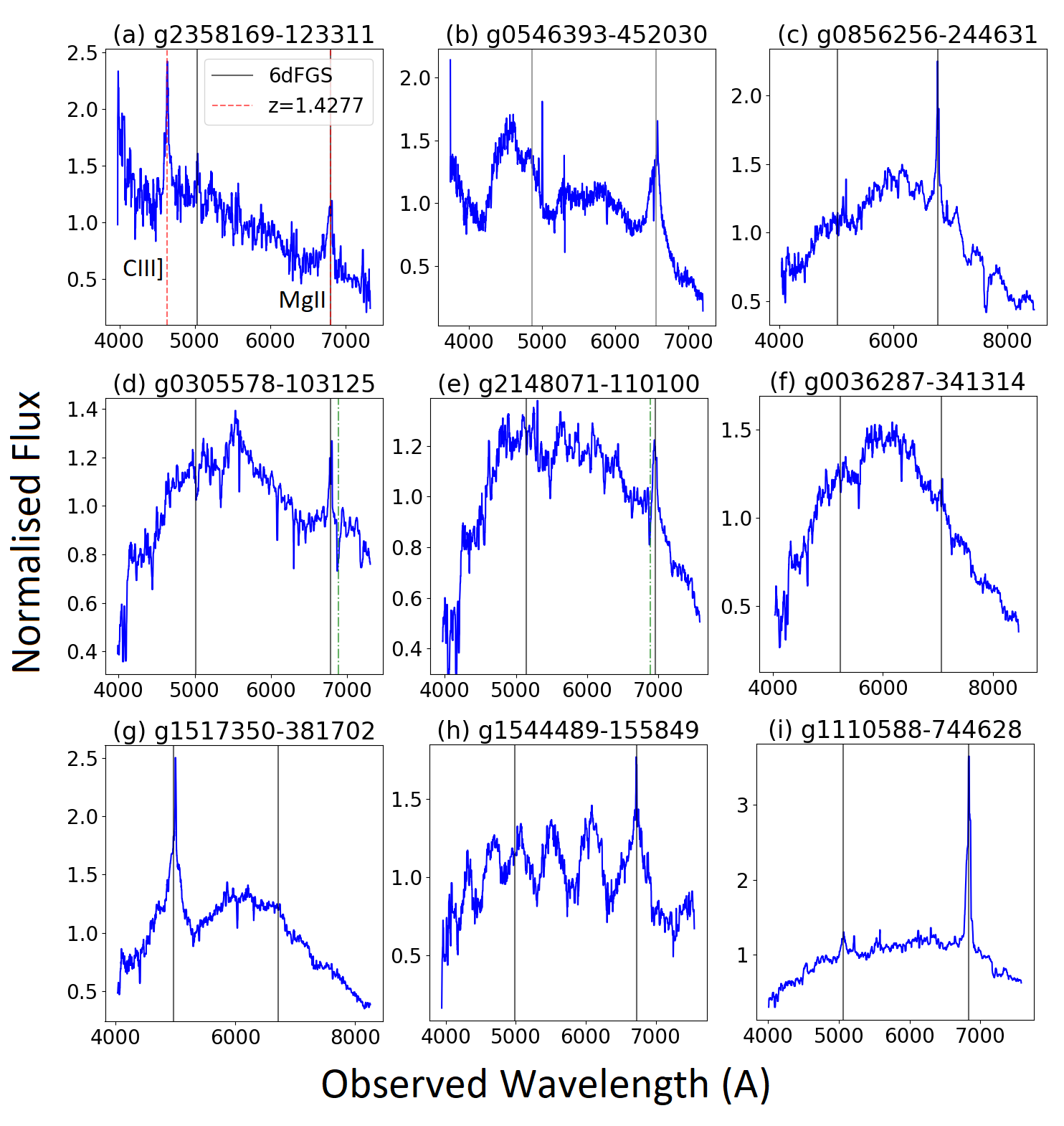}
    \caption{Figures showing examples of problematic cases when comparing our catalogue to 2MAGN. In each, the x-axis is observed wavelength, y-axis is normalised flux, and vertical grey solid lines are indicating H$\beta$ and H$\alpha$ at the 6dFGS redshift. For plot (a), the red vertical dashed lines indicate the labelled emission lines at the corrected redshift. For plots (d, e), the green dashed-dotted lines indicate the B-band telluric absorption.}
    \label{fig:chp:6dfagn:chenex_bad}
\end{figure*}

\begin{figure}
    \centering
    \includegraphics[width=0.42\textwidth]{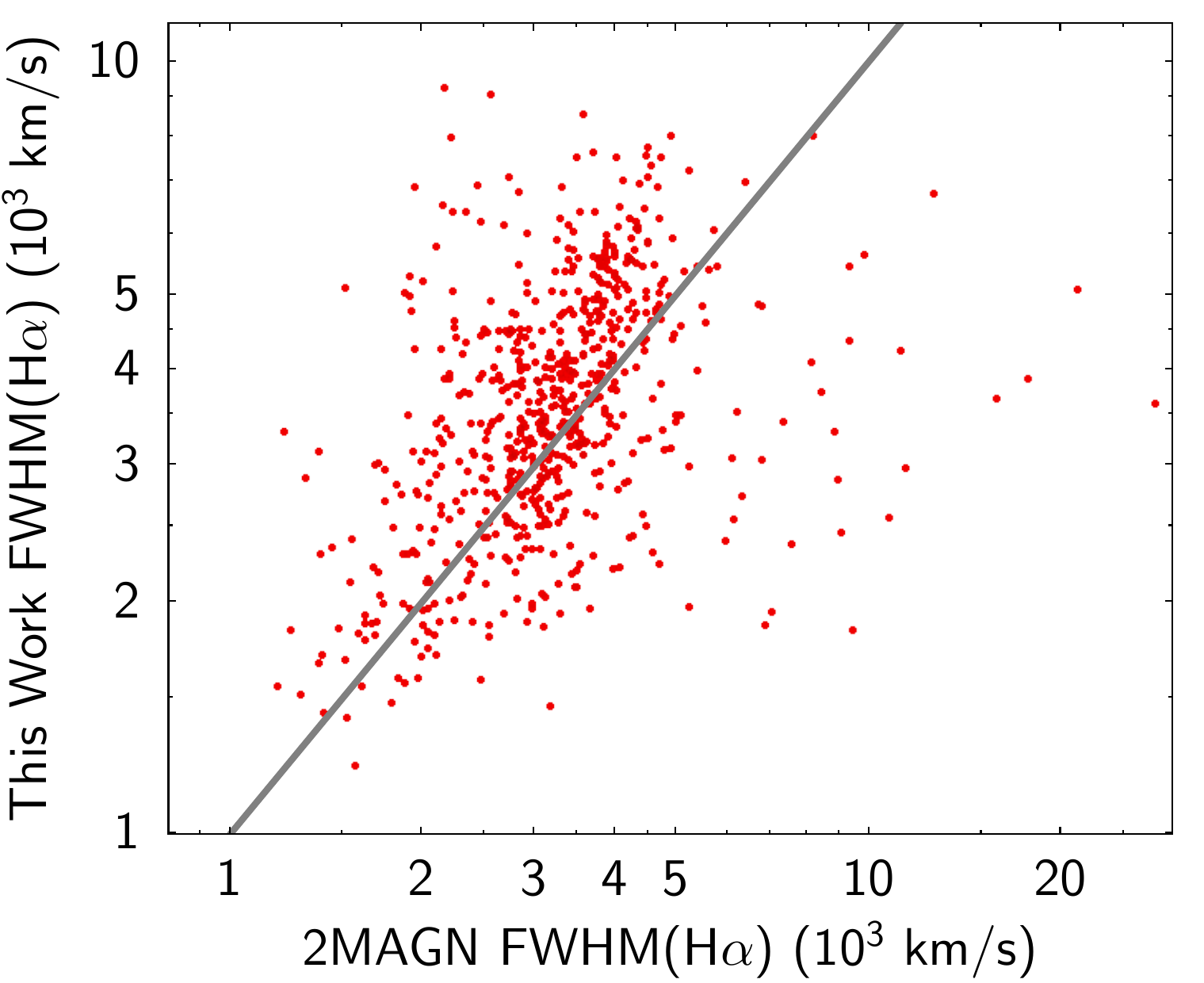}
    \caption{Figure plotting the FWHM values of H$\alpha$ BEL for AGN in our catalogue and in 2MAGN. The red line here is $y=x$ linear line.}
    \label{fig:chen-fwhm}
\end{figure}

\begin{figure*}
    \centering
    \includegraphics[width=0.8\textwidth]{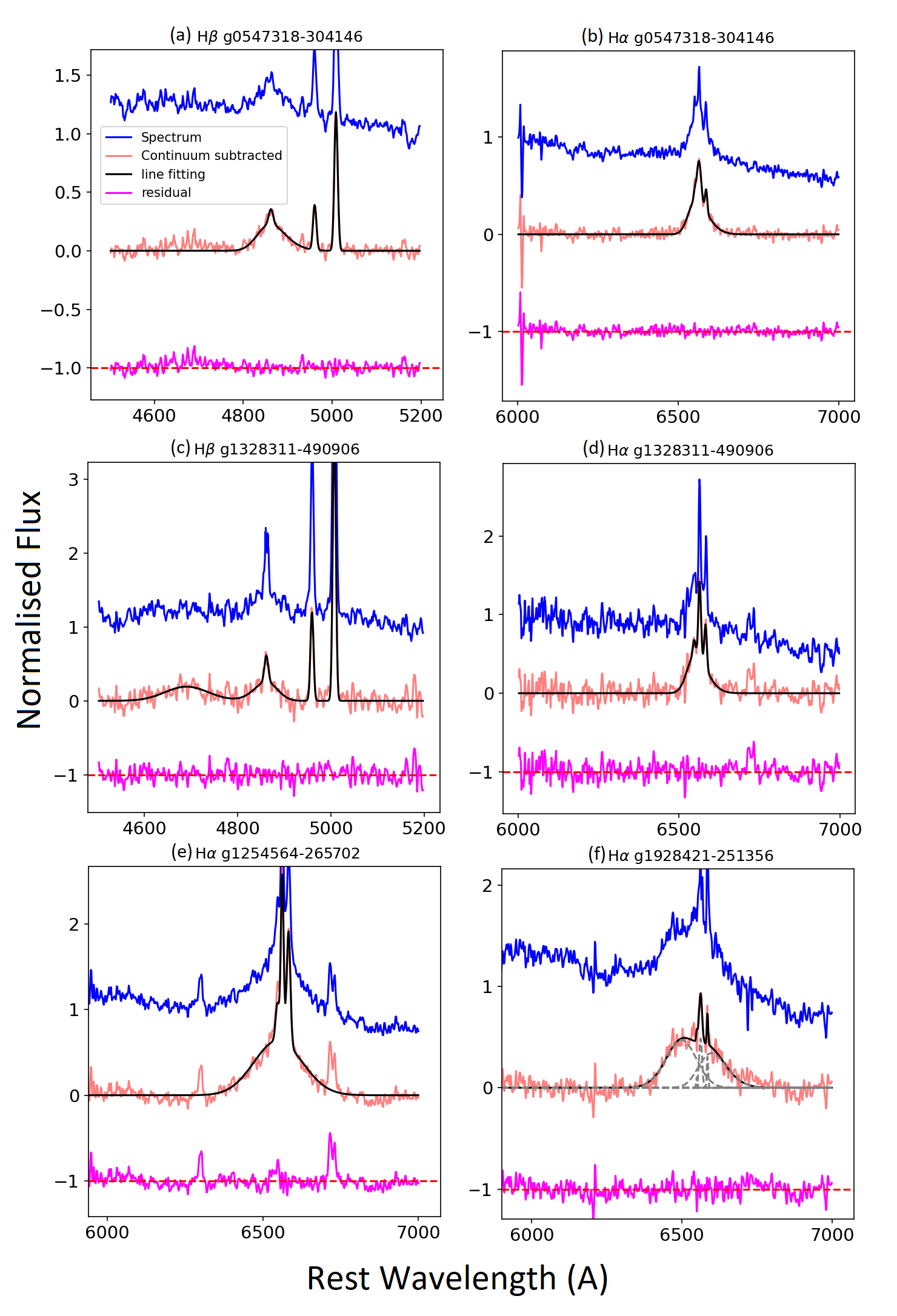}
    \caption{Figures presenting examples of our line fitting for AGN with large discrepancy in FWHM of H$\alpha$ BEL when compared to 2MAGN. In all of these, the x-axis are in rest wavelength, the y-axis are normalised flux, blue line is the original spectrum, pink line is continuum subtracted, black line is the line fitted model, magenta line is the residual (set to mean of -1), and the horizontal red dashed line indicates the zero-level for residual at -1.}
    \label{fig:chp:6dfagn:chenex-fwhm}
\end{figure*}

Therefore, we believe that our BEL AGN list outperforms 2MAGN in several ways. Our catalogue extends the sample beyond $z>0.13$ and exhaustively covers the 6dFGS sample by including sources with lower luminosity and magnitude. We have identified cases with incorrect 6dFGS redshifts, flagged problematic cases, and provided more reliable emission line measurements. Furthermore, our list contains no contaminants, as we have visually inspected every single spectrum and identified any instances of cross-talk. However, we may be missing weak Type-1.9 AGN that our algorithm and tools cannot detect. It is in our interest to recover them as part of a Type-2 AGN-only catalogue in the near future. 
}

\section{Conclusion}
This paper describes the construction of the {\refbf most} complete list of BEL AGN within 6dFGS {\refbf to date. The catalogue is motivated by a need for a hemispheric list of Southern, low-luminosity AGN.} 

In comparison to {\refbf previous attempts at finding optical AGN} in 6dFGS, our catalogue is:
\begin{itemize}
    \item {\refbf Extensive. Starting from a parent sample of 136\,304 spectra with good 6dFGS quality flags, 66\,848 emission line galaxies were selected. 1\,575 of these were already classified in MILLIQUAS, so their line fitting measurements were turned into selection cuts to select BEL AGN. We also selectively inspect discarded spectra to ensure no visually obvious BEL AGN was lost. Our catalogue extends the previously known samples by including dimmer objects across all redshifts. Our final catalogue contains 2\,515 BEL AGN with 1\,674 at $z<0.5$, while 891 are new to the literature.}
    \item Contains no contaminants. Every entry was visually inspected and manually line fitted to ensure that a broad line exist for either H$\alpha$, H$\beta$, or Mg{\sc ii}. {\refbf We have also noted possible sources of contamination arising from fibre cross-talk, eliminating as manycases as we could identify.}
    \item {\refbf Accounts for known 6dFGS problems. Problematic spectra, such as those heavily impacted by the B-band telluric absorption, are flagged. These are considered not fit for use outside of AGN visual identification and redshift estimation.} 
\end{itemize}

{\refbf 
In addition, 39 sources, where 6dFGS reported an incorrect redshift, are flagged in our catalogue with the updated redshifts. Luminosities for the H$\beta$ BEL, [O{\sc iii}]$\lambda5007$, and Mg{\sc ii} are also provided after calibrating the spectra using aperture photometry from the SkyMapper Southern Survey.
}

\section*{Acknowledgement}
We thank the referee for helpful comments in finalising the manuscript. We also thank Katie Auchettl from the University of Melbourne for polishing the draft.

\section*{Data Availability}
All 6dFGS data can be downloaded from their official website \url{http://www-wfau.roe.ac.uk/6dFGS/}. MILLIQUAS catalogue are accessed through TOPCAT under the domain \url{https://heasarc.gsfc.nasa.gov/W3Browse/all/milliquas.html}. BASS data are publicly available from \url{https://www.bass-survey.com/}. The BEL AGN list produced by this work will be uploaded to cds.

%%%%%%%%%%%%%%%%%%%% REFERENCES %%%%%%%%%%%%%%%%%%

% The best way to enter references is to use BibTeX:

\bibliographystyle{mnras}
\bibliography{example} % if your bibtex file is called example.bib

% Don't change these lines
\bsp	% typesetting comment
\label{lastpage}
\end{document}